\documentclass[useAMS,usenatbib]{mn2e}
\pdfoutput=1
\usepackage[utf8]{inputenc}
\usepackage{float}
\usepackage[pdftex]{graphicx}
\usepackage{epstopdf}
\usepackage{color}
\usepackage{soul}
\usepackage{colortbl}
\usepackage{multirow}
\usepackage{setspace}
\usepackage{verbatim}
\usepackage{rotating}  
\usepackage{lscape}
\usepackage{booktabs}
\usepackage{multicol}
\usepackage{bm}  
\RequirePackage{lineno}  
\title[Properties of the GH\,{\normalsize \it II} regions in NGC\,5430]{Properties of the giant H\,{\Large\bf II} regions and bar in the nearby\\ spiral galaxy NGC\,5430}
\author[\'E. Bri\`ere et al.]{\'E. Bri\`ere$^{1,2}$\thanks{E-mail: elaine.briere@queensu.ca (\'EB)}, S. Cantin$^{3}$, and K. Spekkens$^{2}$\\ 
$^{1}$Department of Physics, Engineering Physics \& Astronomy, Queens University, Kingston, Ontario K7L 3N6, Canada\\
$^{2}$Department of Physics, Royal Military College of Canada, PO Box 17000, Station Forces, Kingston, Ontario K7K 7B4, Canada\\
$^{3}$Département de physique, de génie physique et d'optique, and Centre de recherche en astrophysique du Québec (CRAQ),\\ \hspace{0.02cm}
Université Laval, Québec, QC G1K 7P4, Canada}
\def\jnl@style{\it}
\def\aaref@jnl#1{{\jnl@style#1}}

\def\aaref@jnl#1{{\jnl@style#1}}

\def\aj{\aaref@jnl{AJ}}                   
\def\araa{\aaref@jnl{ARA\&A}}             
\def\apj{\aaref@jnl{ApJ}}                 
\def\apjl{\aaref@jnl{ApJ}}                
\def\apjs{\aaref@jnl{ApJS}}               
\def\ao{\aaref@jnl{Appl.~Opt.}}           
\def\apss{\aaref@jnl{Ap\&SS}}             
\def\aap{\aaref@jnl{A\&A}}                
\def\aapr{\aaref@jnl{A\&A~Rev.}}          
\def\aaps{\aaref@jnl{A\&AS}}              
\def\azh{\aaref@jnl{AZh}}                 
\def\baas{\aaref@jnl{BAAS}}               
\def\jrasc{\aaref@jnl{JRASC}}             
\def\memras{\aaref@jnl{MmRAS}}            
\def\mnras{\aaref@jnl{MNRAS}}             
\def\pra{\aaref@jnl{Phys.~Rev.~A}}        
\def\prb{\aaref@jnl{Phys.~Rev.~B}}        
\def\prc{\aaref@jnl{Phys.~Rev.~C}}        
\def\prd{\aaref@jnl{Phys.~Rev.~D}}        
\def\pre{\aaref@jnl{Phys.~Rev.~E}}        
\def\prl{\aaref@jnl{Phys.~Rev.~Lett.}}    
\def\pasp{\aaref@jnl{PASP}}               
\def\pasj{\aaref@jnl{PASJ}}               
\def\qjras{\aaref@jnl{QJRAS}}             
\def\skytel{\aaref@jnl{S\&T}}             
\def\solphys{\aaref@jnl{Sol.~Phys.}}      
\def\sovast{\aaref@jnl{Soviet~Ast.}}      
\def\ssr{\aaref@jnl{Space~Sci.~Rev.}}     
\def\zap{\aaref@jnl{ZAp}}                 
\def\nat{\aaref@jnl{Nature}}              
\def\iaucirc{\aaref@jnl{IAU~Circ.}}       
\def\aplett{\aaref@jnl{Astrophys.~Lett.}} 
\def\apspr{\aaref@jnl{Astrophys.~Space~Phys.~Res.}}
\def\bain{\aaref@jnl{Bull.~Astron.~Inst.~Netherlands}} 
\def\fcp{\aaref@jnl{Fund.~Cosmic~Phys.}}  
\def\gca{\aaref@jnl{Geochim.~Cosmochim.~Acta}}   
\def\grl{\aaref@jnl{Geophys.~Res.~Lett.}} 
\def\jcp{\aaref@jnl{J.~Chem.~Phys.}}      
\def\jgr{\aaref@jnl{J.~Geophys.~Res.}}    
\def\jqsrt{\aaref@jnl{J.~Quant.~Spec.~Radiat.~Transf.}}
\def\memsai{\aaref@jnl{Mem.~Soc.~Astron.~Italiana}}
\def\nphysa{\aaref@jnl{Nucl.~Phys.~A}}   
\def\physrep{\aaref@jnl{Phys.~Rep.}}   
\def\physscr{\aaref@jnl{Phys.~Scr}}   
\def\planss{\aaref@jnl{Planet.~Space~Sci.}}   
\def\procspie{\aaref@jnl{Proc.~SPIE}}   

\begin{document}
\date{Accepted 2012 June 7. Received 2012 May 10; in original form 2011 October 11}
\pagerange{\pageref{firstpage}--\pageref{lastpage}} \pubyear{2012}

\label{firstpage}

\maketitle
\begin{abstract}
In order to better understand the impact of the bar on the evolution of spiral galaxies, we measure the properties of giant H\,{\sc ii} regions and the bar in the SB(s)b galaxy NGC\,5430. 
We use two complementary data sets, both obtained at the Observatoire du Mont-M\'egantic: a hyperspectral data cube from the imaging Fourier transform spectrograph SpIOMM, and high-resolution spectra across the bar from a long-slit spectrograph. 
We flux-calibrate SpIOMM spectra for the first time, and produce H$\alpha$ and [N\,{\sc ii}]$\lambda$6584\,\r{A} intensity maps from which we identify 51 giant H\,{\sc ii} regions in the spiral arms and bar. 
We evaluate the type of activity, the oxygen abundance and the age of the young populations contained in these giant H\,{\sc ii} regions and in the bar. 
Thus, we confirm that NGC\,5430 does not harbour a strong AGN, and that its Wolf-Rayet knot shows a pure H {\sc ii} region nature.
We find no variation in abundance or age between the bar and spiral arms, nor as a function of galactocentric radius. 
These results are consistent with the hypothesis that a chemical mixing mechanism is at work in the galaxy's disc to  flatten the oxygen abundance gradient.
Using the {\sc starburst99} model, we estimate the ages of the young populations, and again find no variations in age between the bar and the arms or as a function of radius. 
Instead, we find evidence for two galaxy-wide waves of star formation, about 7.1\,Myr and 10.5\,Myr ago. 
While the bar in NGC\,5430 is an obvious candidate to trigger these two episodes, it is not clear how the bar could induce widespread star formation on such a short time-scale. 
\end{abstract}

\section[]{Introduction}


 Observations reveal that two thirds of spiral galaxies are barred \citep{1963ApJS....8...31D,2000AJ....119..536E,2007ApJ...659.1176M}. 
The formation mechanisms and the impact of this structure on the disc are not fully understood. 
However, some observations and simulations suggest that bars are temporary or cyclical features in spiral galaxy evolution \citep{2002ASPC..275..157D}.
This structure would facilitate the distribution of chemical elements in the disc and toward the centre \citep{1996ApJ...462..114N,2011A&A...529A..45V}.
Thereby, a bar would tend to flatten or even erase an abundance gradient in the disc \citep{1981A&A...101..377A,1992MNRAS.259..121V,1994ApJ...420...87Z,1994ApJ...430L.105F,1994ApJ...424..599M,1995ApJ...445..161M,1996ASPC...91...63R, 1997MNRAS.288..715R,1999ApJ...516...62D,2010MNRAS.406.1094F}, probably as a consequence of the non-circular motions induced by this kind of structure \citep{2011arXiv1101.1771V}. 
Moreover, the presence of a bar influences the location of massive star formation in the galaxy \citep{2007A&A...465L...1W}, but not the star formation rate \citep{1994mtia.conf..131K}. 

One way to investigate the impact of a bar in a single galaxy is to measure the properties of H\,{\sc ii} regions in and beyond the bar, such as their position in the disc, O/H abundance and young population age \citep{1983ApJ...267..563H,1983ApJ...272...54K,1994ApJ...420...87Z,1994mtia.conf..131K}.
However, for distant galaxies, individual H\,{\sc ii} regions cannot be resolved. 
For these systems, we instead consider the properties of giant H\,{\sc ii} regions, i.e. the aggregate of several H\,{\sc ii} regions.

In this paper, we present a study of the activity, abundances and young population ages in the arms and bar of the spiral galaxy NGC\,5430.
We use optical spectra obtained with the imaging Fourier transform spectrograph SpIOMM and a long-slit spectrograph at the Observatoire du Mont-M\'egantic.
The SpIOMM imaging capacity provides a spectrum for each pixel across the 12$\arcmin$ circular field-of-view, albeit with lower throughput than a long-slit spectrograph. 
These two data sets are therefore complementary, and we use the sensitivity of the long-slit spectrograph to corroborate the 2D results obtained with SpIOMM.

\begin{figure}\begin{minipage}{1\linewidth}\centering
\includegraphics[trim = 0.1cm 0.3cm 0.1cm 0.1cm, clip, width=1\textwidth]{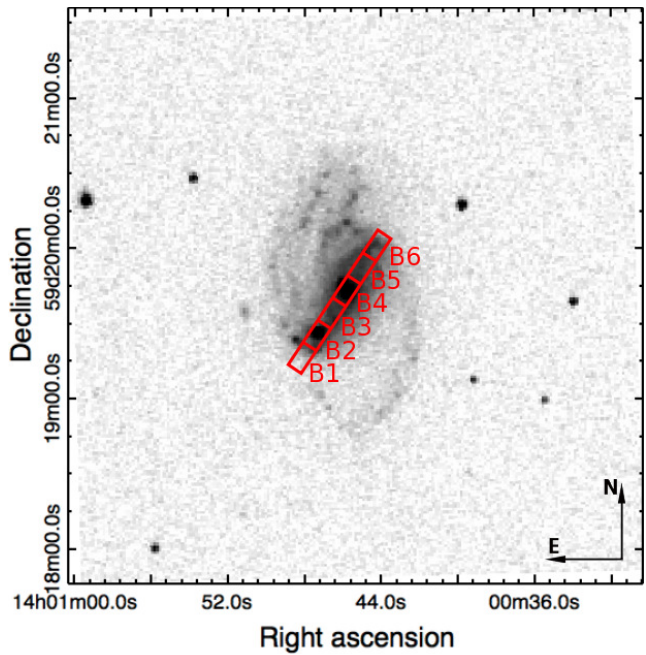}
\caption{Image of NGC\,5430 obtained by averaging SpIOMM exposures in the $r\arcmin$ filter before the Fourier transform is taken. The red boxes show the six 10.4$\arcsec$ x 4$\arcsec$ zones along the bar from which long-slit spectra were extracted.\label{fig:n5430}}\end{minipage}\end{figure}

NGC\,5430 is a SB(s)b galaxy (RA 14$^\rmn{h}$00$^\rmn{m}$45\fs6, Dec. +59$\degr$19$^\rmn{m}$42$^\rmn{s}$, J2000) located at a distance of 42$\pm$3~Mpc, using a redshift of $cz$=3061$\pm$9~km\,s$^{-1}$ \citep{1996ApJ...473..576F} and $H_0$\,=\,73.0~km\,s$^{-1}$\,Mpc$^{-1}$. 
Two regions in the galaxy exhibit strong emission as shown in Fig.~\ref{fig:n5430}.
The brightest region is located 22$\arcsec$ from the galaxy nucleus (within B2 in Fig.~\ref{fig:n5430}) and is identified as a Wolf-Rayet (WR) knot \citep{1982PASP...94..765K,1987A&A...172...43K,2004MNRAS.355..728F}.
The second is the central region (within B4 in Fig.~\ref{fig:n5430}), which shows active star formation in two knots on a nuclear ring \citep{1983ApJ...268..602B,1997A&A...324...41C,Cantin2004,Cantin2010}.
This galaxy has an inclination of 50$\pm$5$\degr$ and a position angle of 177$\pm$16$\degr$ \citep{2008MNRAS.388..500E}. 
NGC\,5430 is one of the first extragalactic sources studied with SpIOMM. 

Using the spatial and spectral information from the long-slit spectrograph and SpIOMM, we evaluate the type of activity (H\,{\sc ii}, LINER, composite, etc. $-$ \citealt*{1981PASP...93....5B}; \citealt{2006MNRAS.372..961K}), the abundance \citep{2002ApJS..142...35K} and the young population age ({\sc starburst99}, \citealt{1999ApJS..123....3L}) of the NGC\,5430 bar and giant H\,{\sc ii} regions.
Then we look at the behavior of these parameters as function of galactocentric radius, in the bar and in the spiral arms.

\begin{figure*}\begin{minipage}{1\linewidth}\centering
\includegraphics[trim = 0.4cm 0.4cm 0.2cm 0.5cm, clip, width=0.88\textwidth]{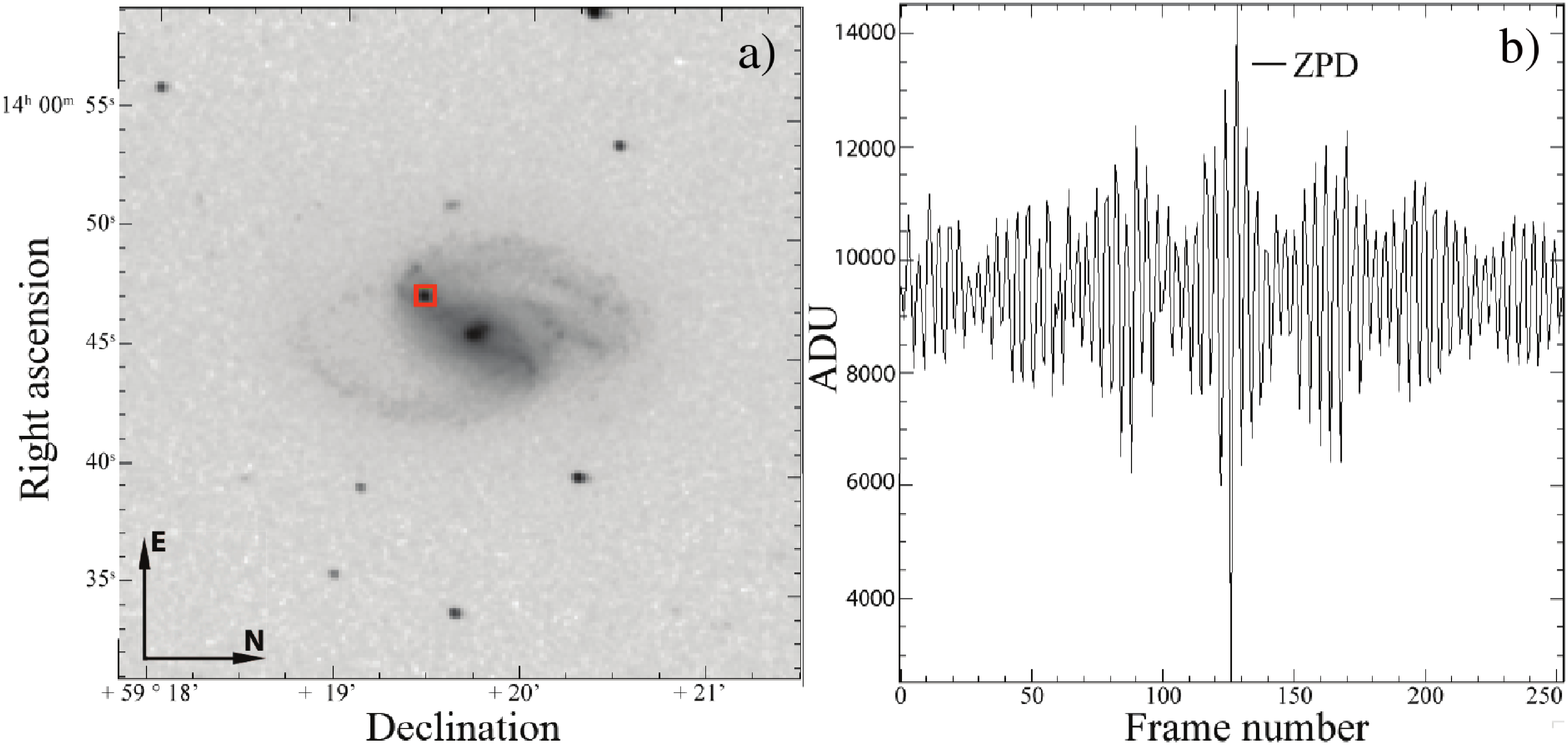}
\includegraphics[trim = 0.4cm 0.4cm 0.2cm 0cm, clip, width=0.88\textwidth]{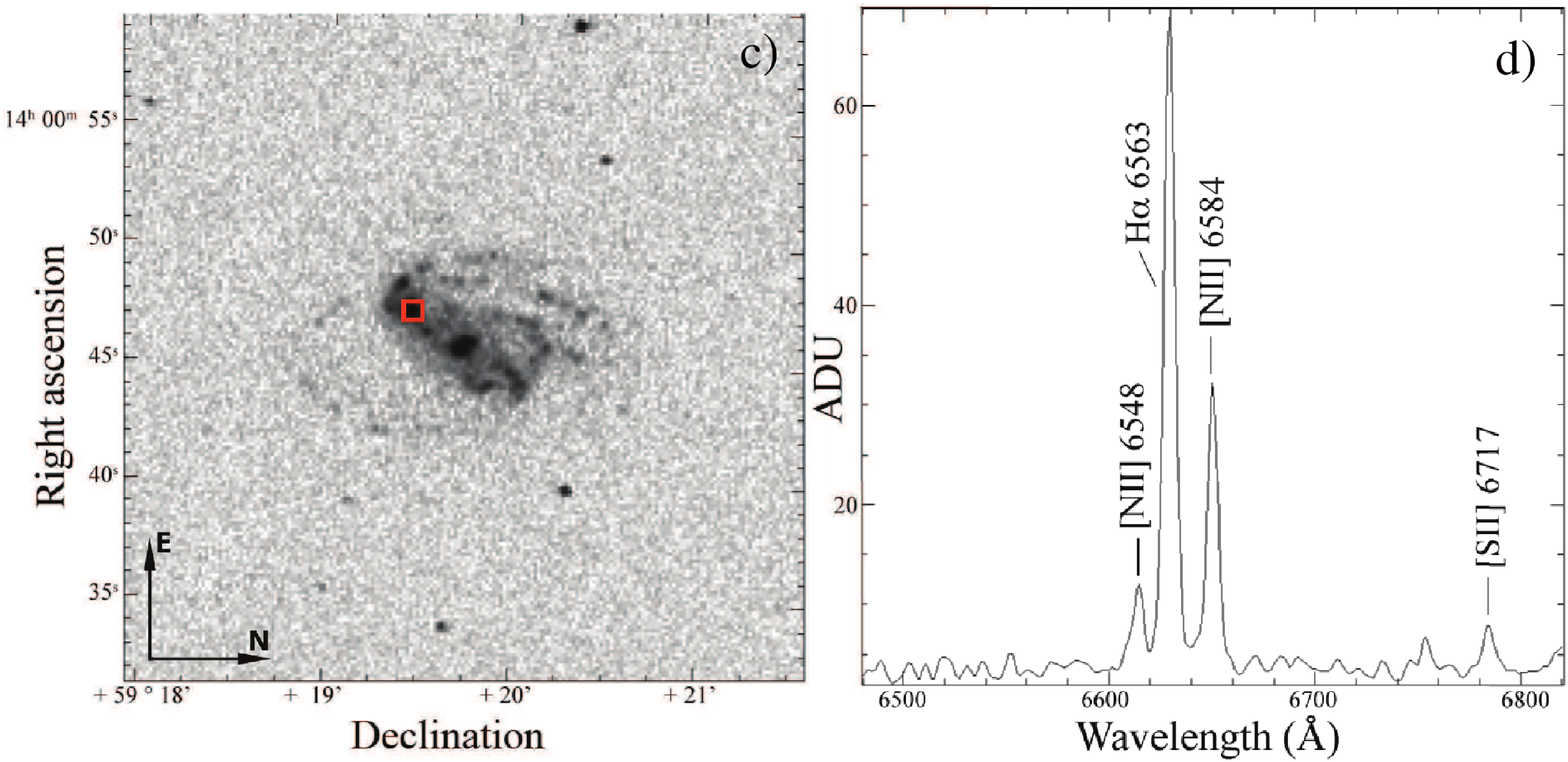}
  \caption{Example of data obtained with SpIOMM. a) Image of NGC\,5430, before the FFT. The red box shows the region in which the interferograms are summed to produce the pulse in b), which is characteristic for bright giant H\,{\sc ii} regions. Additionally, the zero path difference (ZPD) peak can be easily identified. c) H$\alpha$ intensity map of NGC\,5430, obtained by integrating the hyperspectral cube in frequency after the FFT. The red box shows the same region as in a). d) Spectrum of the boxed region in c), with spectral features labelled. \label{tf} }\end{minipage}\end{figure*}

Details about the observations, the data reduction and the analysis procedure, including how we evaluate the reddening and the giant H\,{\sc ii} region boundaries, are presented in Section~2 of this paper. 
Our results are found in Section~3 and compared with others found in the literature. 
Finally, Sections~4 and 5 contain a discussion and a summary of our conclusions, respectively.

\setcounter{figure}{2}
\begin{figure*}\begin{center}
\includegraphics[trim = 0cm 0cm 2.6cm 0cm, clip, width=0.481\textwidth]{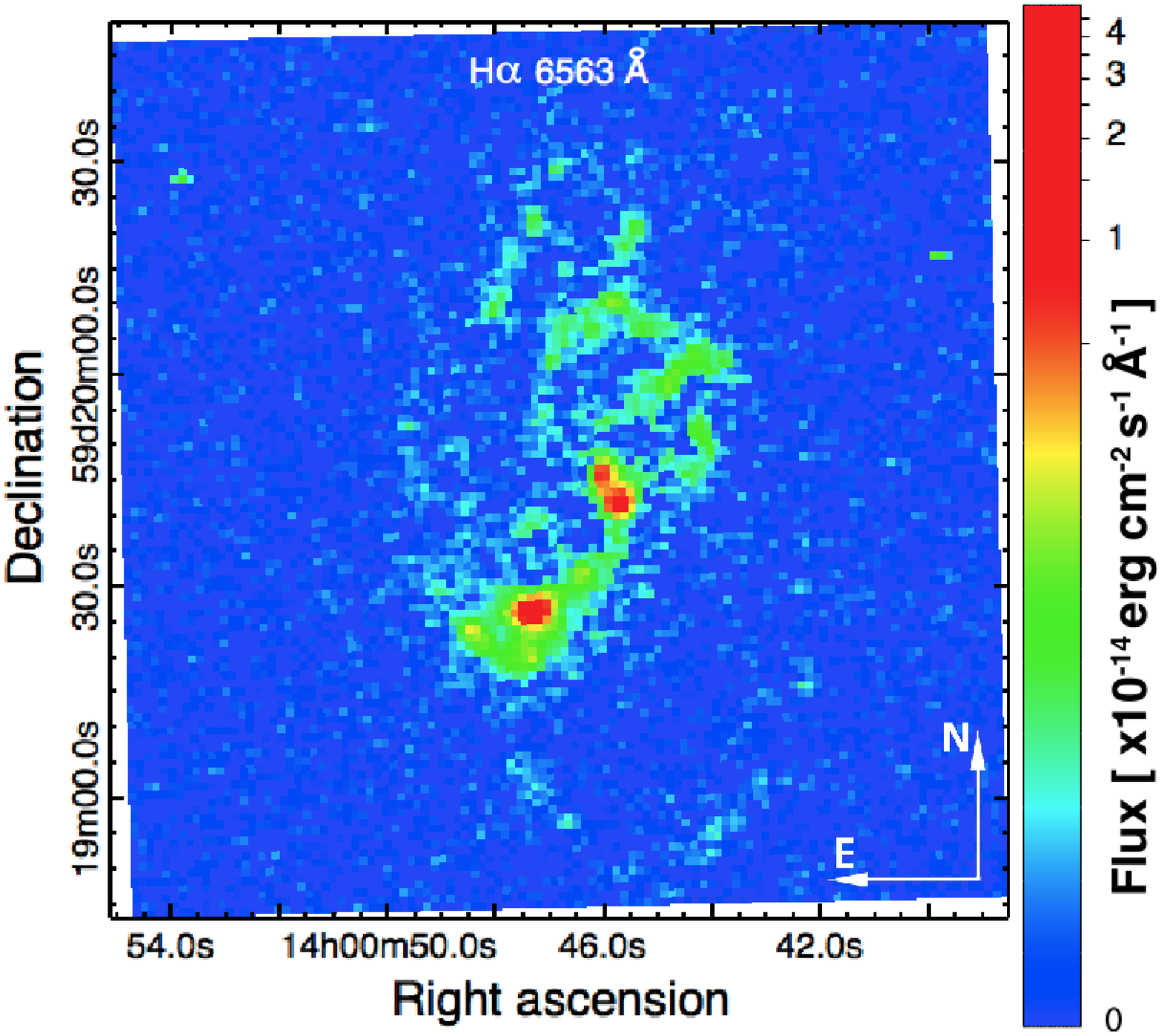}
\includegraphics[trim = 1.9cm 0cm 0cm 0cm, clip, width=0.499\textwidth]{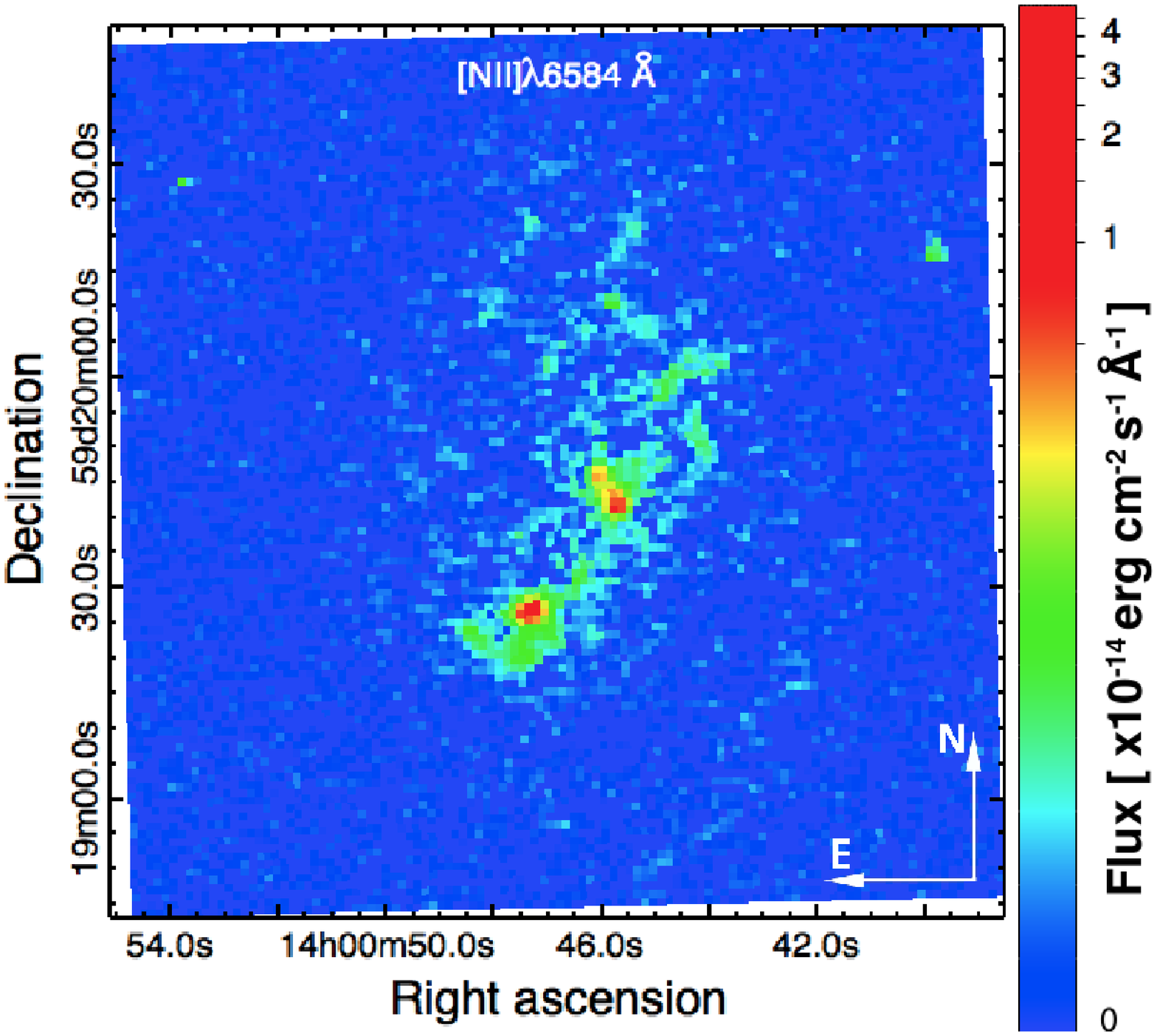}
\caption{SpIOMM intensity maps of H$\alpha$ and [N\,{\sc ii}]\,$\lambda$6584 emission lines. The colour scale is logarithmic. \label{fig:maps_int}}
\end{center}\end{figure*}

\section[]{Observations and Data Reduction}

The two data sets used for our study were obtained with the 1.6-meter Ritchey-Chrétien, Boller \& Chivens telescope at the Observatoire du Mont-M\'egantic between March 2006 and May 2008. The sky conditions were good to excellent during the three observing runs, with a typical seeing of 1.3$\arcsec$.

\subsection[]{SpIOMM Observations}

A hyper-spectral cube centered on NGC\,5430 was obtained with the imaging Fourier transform spectrograph SpIOMM on 2008 May 5. The specific parameters of the observations and the reduced data cube are presented in Table~\ref{tab:spiommdata}.

SpIOMM is a modified Michelson interferometer that obtains morphological, spectroscopic and kinematic information over a 12$\arcmin$ circular field-of-view in a single observing sequence.
During data acquisition, one of the two mirrors constituting the instrument moves at regular steps of few nanometers.
At each step (or relative phase), a new image is taken. 
Together, these images produce a data cube in right ascension, declination and relative phase that contains an interference pattern, or interferogram, for each pixel of the CCD. 
Fig.~\ref{tf} is an example of a summed interferogram and its associated spectrum for a bright giant H\,{\sc ii} region.
At the zero path difference (ZPD), usually located in the centre of the interferogram, all wavelengths interfere constructively or destructively producing a peak as shown in Fig.~\ref{tf} b). 
This peak contains the information along the line of sight that corresponds to the fast Fourier transform (FFT) of the spectrum. 
Sometimes a pulse can be observed in the interferogram, a sign of strong emission lines in the associated spectrum.  
For further information about the instrument and the observing procedure see \cite{2010AJ....139.2083C} and other related publications \citep{Lavigne2003,2003SPIE.4842..392G,Rochon2006,Grandmont2006,2006SPIE.6269E.135B,Charlebois2008,2008SPIE.7014E.245B,2008SPIE.7014E.246D}.

\begin{table}
 \centering
 \begin{minipage}{80mm}
  \caption{SpIOMM observations and data cube parameters.\label{tab:spiommdata}}
\begin{tabular}{@{}lc@{}}\toprule 
Date							&2008 May 5		\\
Seeing						&$\sim$1.3$\arcsec$		\\
Filter							&$r\arcmin$ (6480-6820\,\r{A})\\ 
Alias order					&16				\\
OPD\footnote{Optical Path Difference (OPD).} sampling distance			&5542 nm			\\
Exposure time per image 	       		&45	s				\\
Number of images		&253		\\
Resulting resolution $\Delta\sigma$	&8.64 cm$^{-1}$	\\
Spectral resolution $\Delta\lambda$ \footnote{Relationship between the resulting and spectral resolution: $\Delta\sigma=\Delta\lambda/\lambda^2$.}  &3.62 to 4.01~\r{A}	\\
Pixel scale				 & 1.1$\arcsec$/pixel 			\\
Total observing time			&3h09m				\\\bottomrule
\end{tabular}\end{minipage}\end{table}

We used the intermediate-band filter $r\arcmin$ (6480-6820\,\r{A)}) to limit the bandwidth to the [N\,{\sc ii}]\,$\lambda\lambda$6548,6584, H$\alpha$\,$\lambda$6563 and [S\,{\sc ii}]\,$\lambda$6716 emission lines.
Note that because of NGC\,5430's redshift the second line of the sulfur doublet, [S\,{\sc ii}]\,$\lambda$6731, falls outside of the filter bandpass. 
Every image in the cube was corrected for CCD readout noise, pixel-to-pixel variations, sky brightness and cloud cover variations.
We also excised the cosmic rays and satellite tracks that were at least three times greater than the interferogram standard deviation. 
The SpIOMM spectra have not been corrected for internal extinction because their bandwidth does not include H$\beta$. 
However, we restrict our analysis using these data to the [N\,{\sc ii}]\,$\lambda$6584/H$\alpha$ line ratio.
These two emission lines are close in wavelength in the red portion of the spectrum, which is less affected by reddening. 
Therefore, the impact of reddening on this ratio should be minimal.

Once all interferograms were corrected, we took an FFT to convert them into spectra and we then applied a wavelength calibration \citep{2003SPIE.4842..392G,Grandmont2006}. 
In addition, a flux calibration has been applied for the first time to SpIOMM data. 
For this, we summed all images in the frequency domain, and we measured the instrumental magnitude of about ten stars throughout the field-of-view. 
Using the relation presented in \cite{1996AJ....111.1748F}, we converted these magnitudes into fluxes and compared them to the fluxes measured by the Sloan Digital Sky Survey \citep{2008ApJS..175..297A}. 
That gave us an average flux ratio $f_{SLOAN}/f_{SpIOMM}$\,=\,6.2$\pm$0.8 by which we multiplied each spectrum. 
Finally, each spectrum was corrected for Milky Way extinction, with $E(B-V)$\,=\,0.015 \citep{1998ApJ...500..525S}, and assigned appropriate spatial coordinates with the {\sc karma} task {\it koords}  \citep{1995ASPC...77..144G,1996ASPC..101...80G} and a Digital Sky Survey image. 

\subsubsection[]{Intensity Maps\label{intensity}}

Exploiting the SpIOMM imaging capacity, we obtained the H$\alpha$ and [N\,{\sc ii}]\,$\lambda$6584 emission line intensity maps presented in Fig.~\ref{fig:maps_int}. 
Each pixel in the map is the sum of the continuum-subtracted emission in that spectral line along the line-of-sight.

As shown in Figs.~\ref{fig:n5430} and \ref{tf}, NGC\,5430 has a strong oval-shaped bar that we identified as a type B in the \cite{1997A&A...326..449M,1999A&A...346..769M} classification scheme.
This means that several H\,{\sc ii} regions are visible in the bar and star formation is occurring in the galaxy's core.
In both intensity maps, the emission is strong in the core, on the leading side of the bar, in the northern arms and in the south-east knot. 
These observations are consistent with the \cite{2005A&A...439..539O} and \cite{1996ASPC..103..175C} isophotal contour maps.
The strongest emission in the galaxy is in a region of diameter $\sim$2$\arcsec$\,=\,0.4\,kpc corresponding to the WR knot \citep{1982PASP...94..765K,1987A&A...172...43K}. 
In the galaxy centre, the emission is particularly intense in two regions located on each side of the core and on a circum-nuclear ring \citep{1997A&A...324...41C,Cantin2010}.
This central ring is not evident in the SpIOMM data due to the limited spatial resolution ($\sim$1.3$\arcsec$).  
Finally, the northern arms exhibit stronger emission than the southern arms. According to \cite{1998MNRAS.301..631R}, this type of asymmetry is widely observed in barred galaxies and related to the strength of this structure. Hence, the stronger the bar, the less symmetric the star formation in the outer disc.
Moreover, the relative strength of emission lines suggests that the brighter parts of the galaxy are affected by thermal processes since the H$\alpha$ emission is stronger than the [N~{\sc ii}] emission. 
However, there is no clear indication of a close companion in our data that could explain this perturbation \citep{1987A&A...172...43K}.
Additionally, \cite{2003ApJ...598..260P} consider NGC\,5430 as an isolated galaxy. 

\begin{figure}\begin{minipage}{1\linewidth}
\begin{center}
\includegraphics[trim = 0cm 0cm 0cm 0cm, clip, width=1\textwidth]{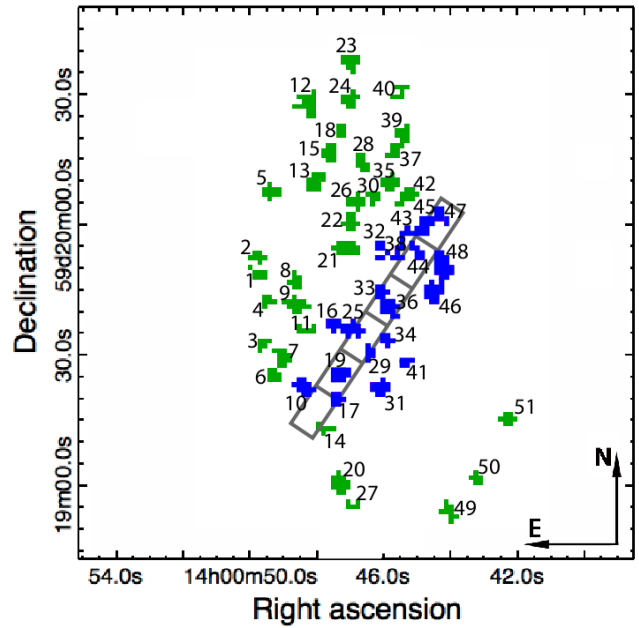}
\caption{Giant H\,{\sc ii} regions identified in SpIOMM intensity maps in Fig.~\ref{fig:maps_int}. Regions in the bar are in blue, and those in the arms are in green. The properties of each numbered H\,{\sc ii} region are given in Table~\ref{tab:result_spiomm}. The grey boxes show the locations and sizes of zones B1-B6 (see Fig.~\ref{fig:n5430}).  \label{maps_HII}}
\end{center}\end{minipage}\end{figure}

\subsubsection[]{Giant H {\sevensize\it II} Region Identification\label{GHIIR}}

\setcounter{figure}{4}
\begin{center}\begin{figure*}
  \includegraphics[trim = 0mm 15mm 8.5mm 0mm, clip, width=0.51\textwidth]{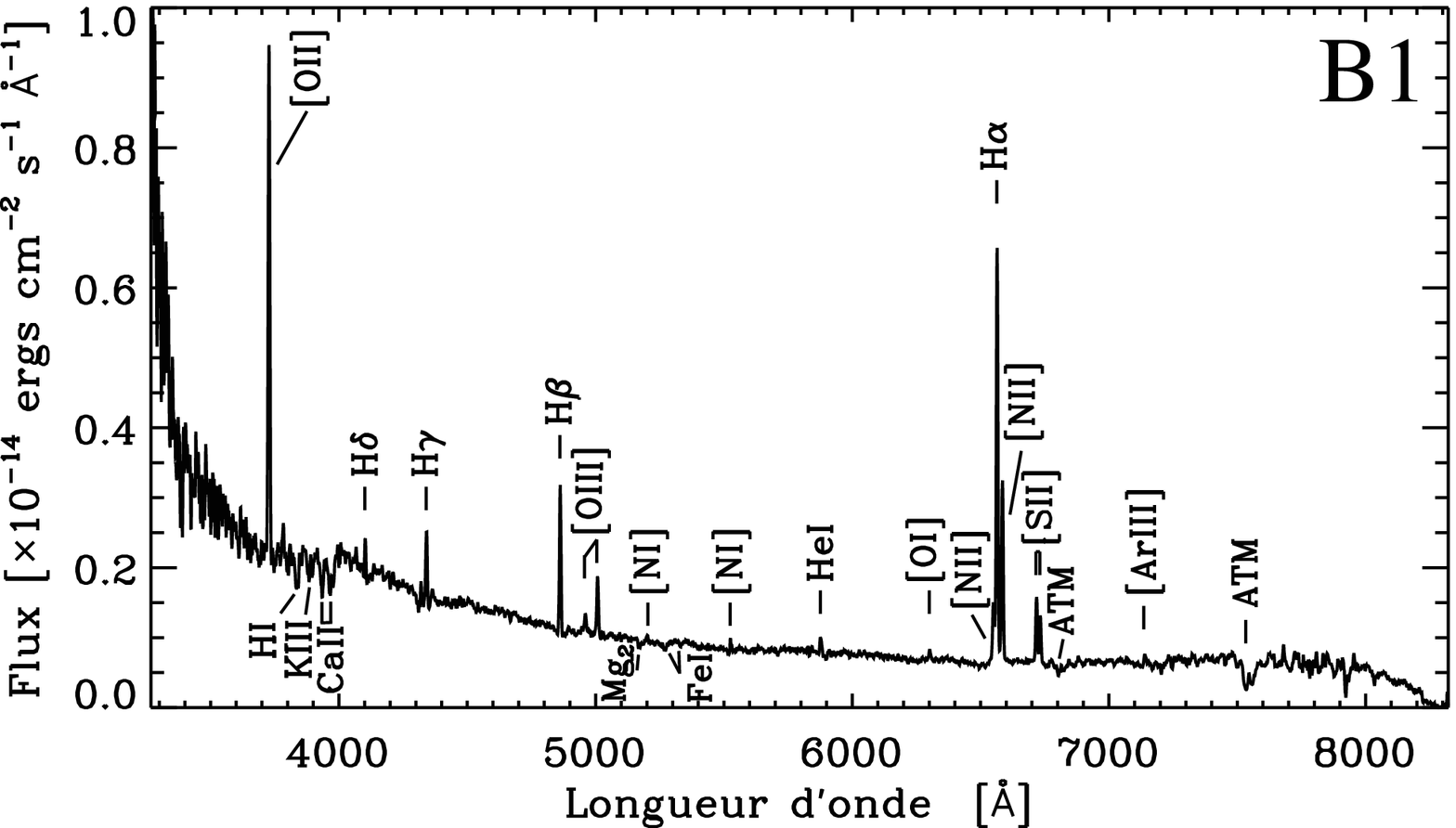}
  \includegraphics[trim = 9mm 15mm 9mm 0mm, clip, width=0.472\textwidth]{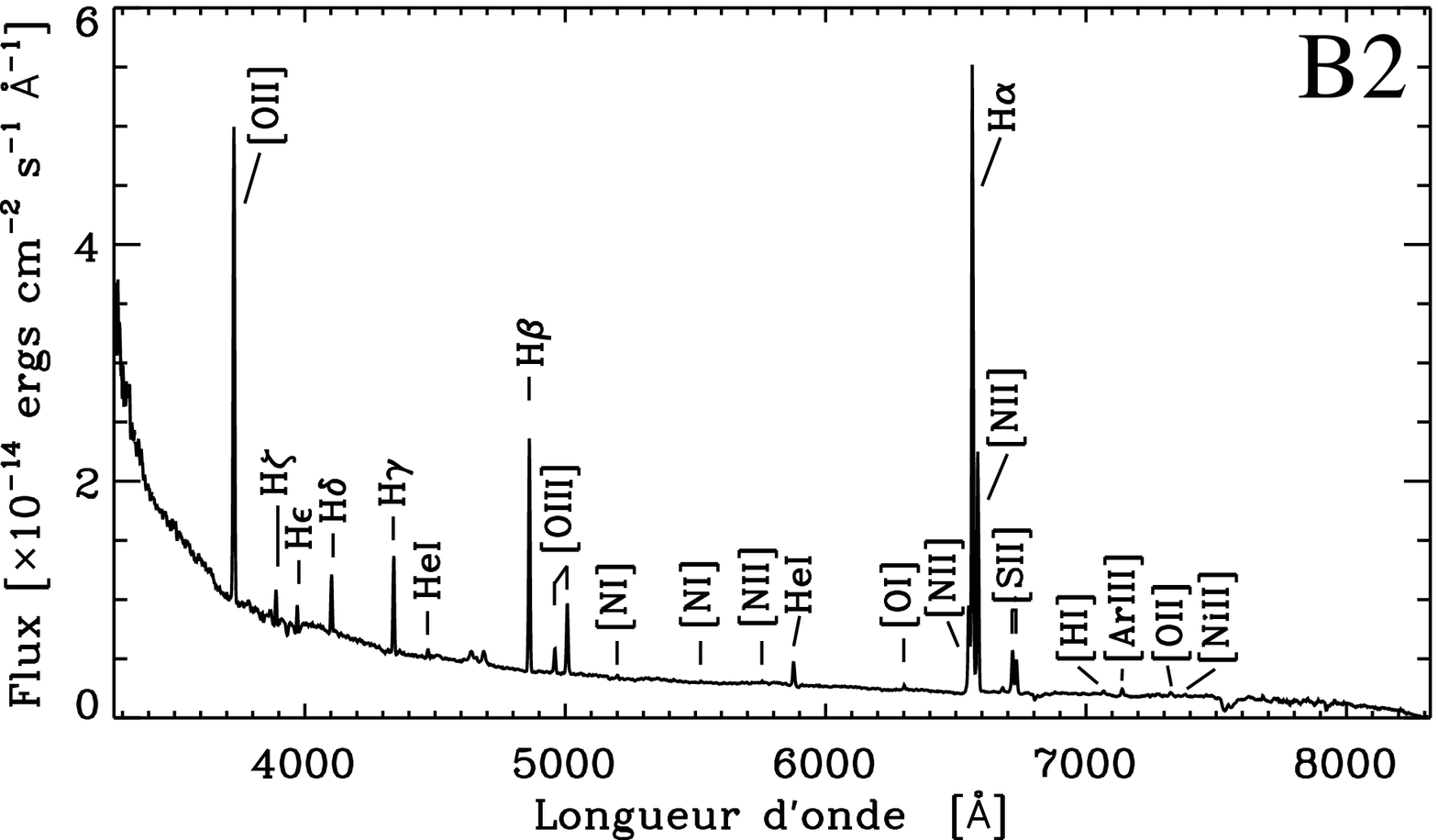}
  \includegraphics[trim = -2.5mm 15mm 9mm 0mm, clip, width=0.51\textwidth]{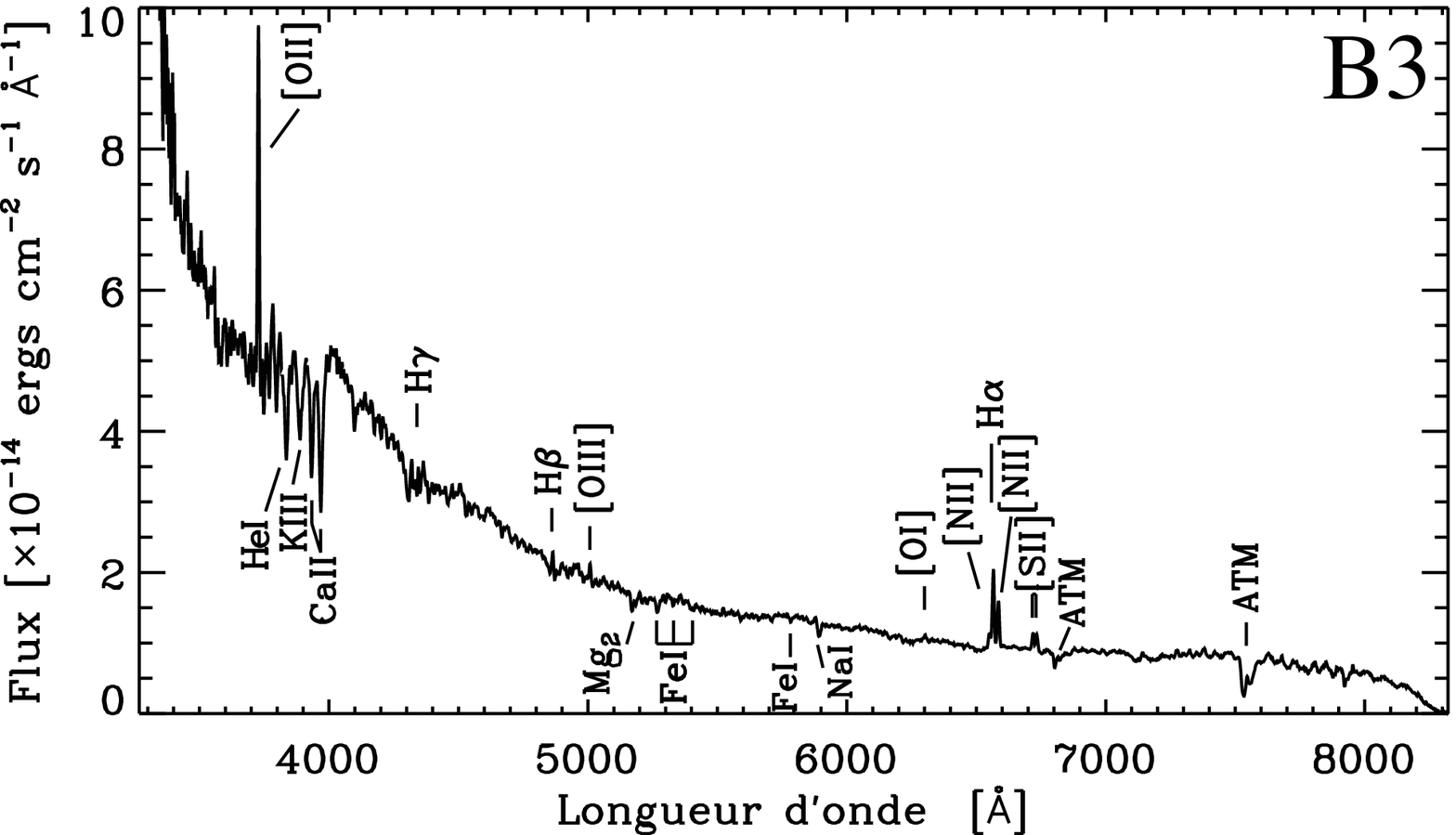}
  \includegraphics[trim = 9mm 15mm 9mm 0mm, clip, width=0.472\textwidth]{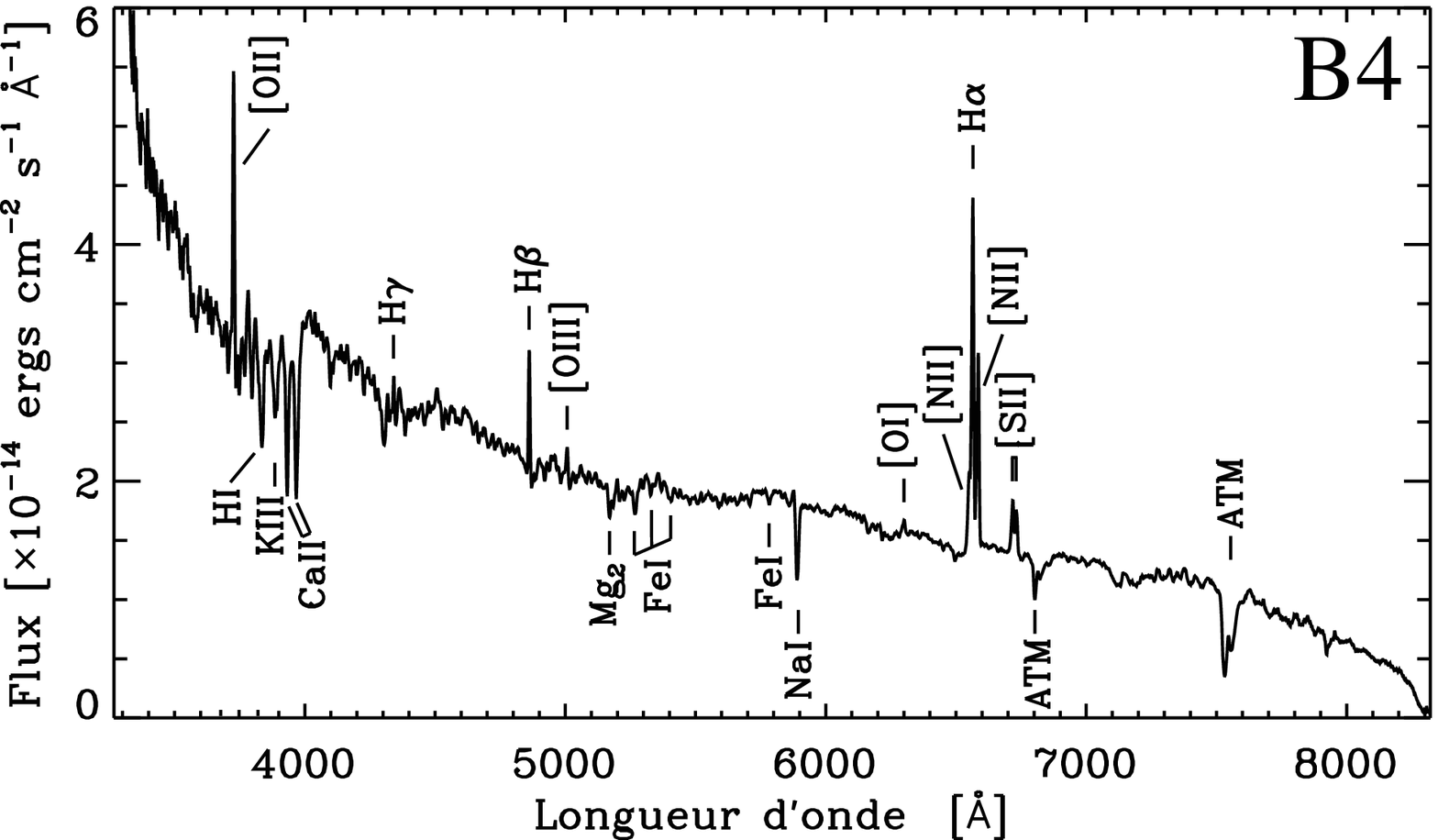}
  \includegraphics[trim =  -2.6mm 0mm 9mm 0mm, clip, width=0.51\textwidth]{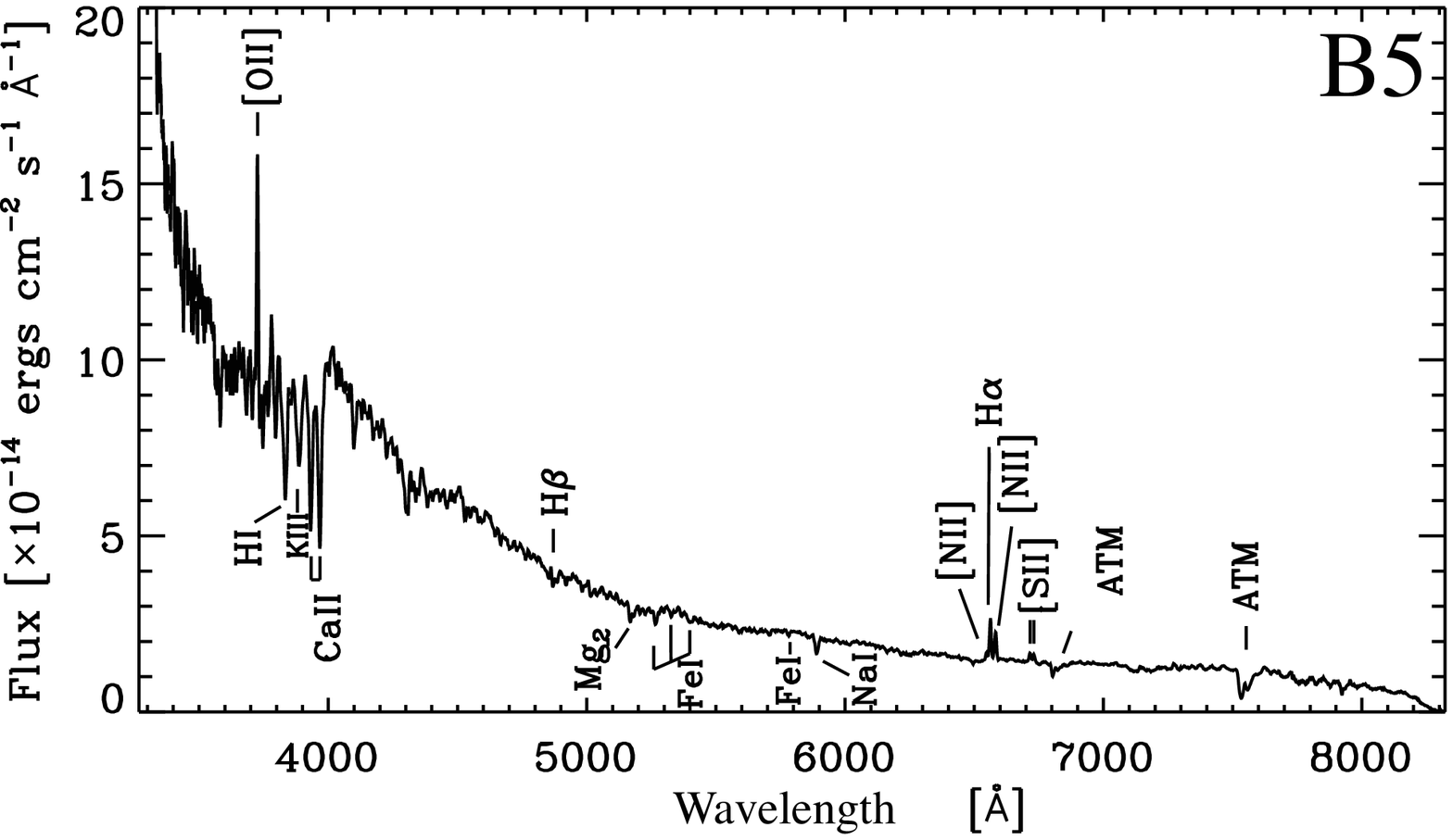}
  \includegraphics[trim =  9mm 0mm 9mm 0mm, clip, width=0.472\textwidth]{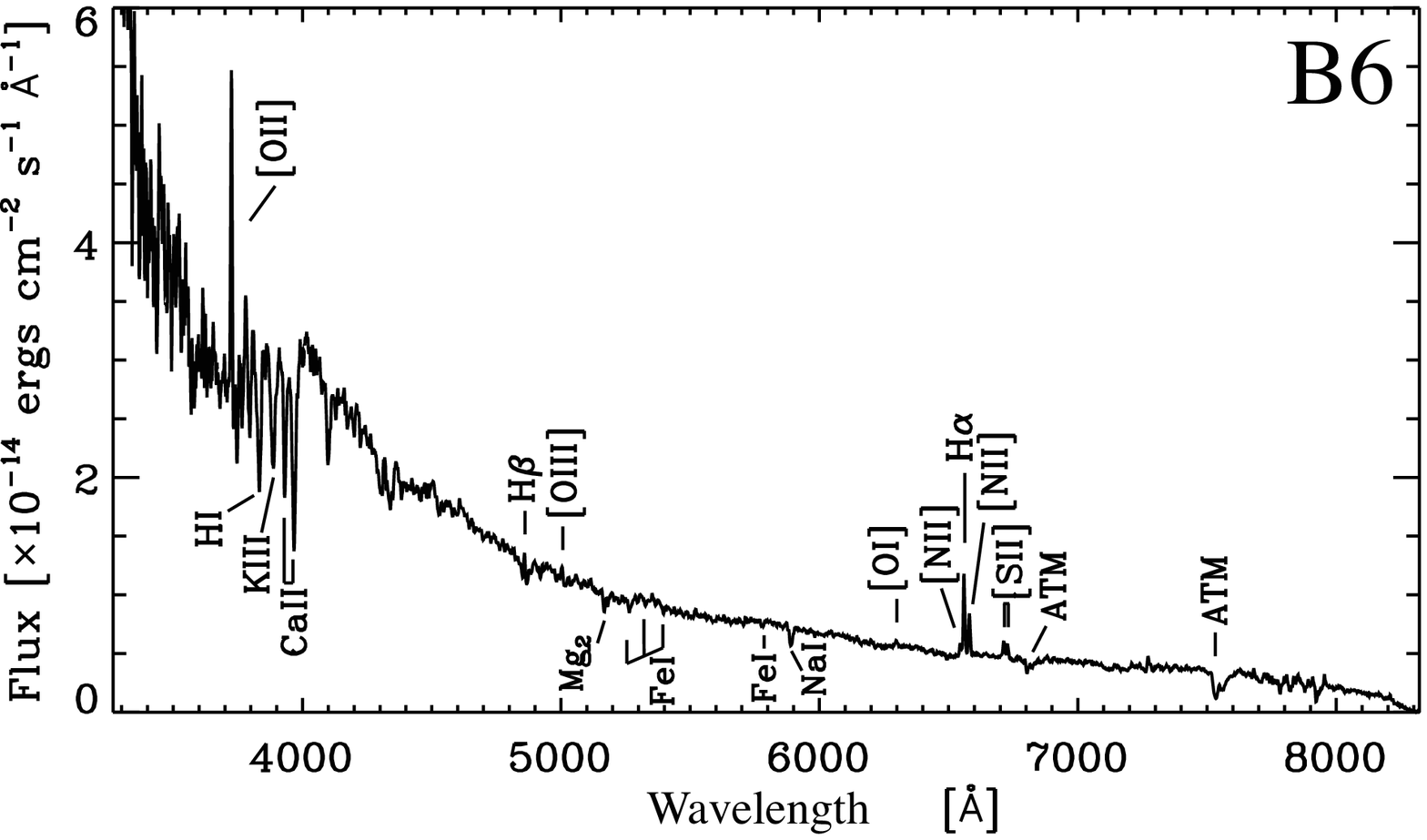}
 \caption{Long-slit spectra across the NGC\,5430 bar, summed over the zones shown in Fig.~\ref{fig:n5430}. In each panel, the detected spectral transitions are labelled, and atmospheric artifacts are indicated by {\it ATM}.
  \label{spectres}}\end{figure*}\end{center}

In order to measure the properties of the giant H\,{\sc ii} regions throughout NGC\,5430, we use the SpIOMM data to identify these regions and to produce a single spectrum for each. 
Using an intensity map summed over both the H$\alpha$ and [N\,{\sc ii}]\,$\lambda$6584 emission lines and the {\sc idl} program {\it H{\sevensize\it II}phot.pro} \citep{2000AJ....120.3070T}, we detected 99 bright regions. 
We then summed the SpIOMM spectra within the identified regions. 
From them, we selected 51 giant H\,{\sc ii} regions with a signal-to-noise S/N\,$\geq$\,5 in the H$\alpha$ line for our analysis.
We separated this sample into regions located in the bar and those located in the arms by visual inspection. 
Fig.~\ref{maps_HII} shows a map of these 51 giant H\,{\sc ii} regions and Table~\ref{tab:result_spiomm} contains their measured properties.

\subsection[]{Long-slit Spectrograph Observations}

With the Perkin-Elmer long-slit spectrograph at the f/8 focus and a 600 lines\,mm$^{-1}$ diffraction grating, we took observations from 2006 March 26-31 along the NGC\,5430 bar that have a higher sensitivity than the SpIOMM observations described above. 
For these 16 exposures of 2700 seconds the slit was wide open (width: 4$\arcsec$, length: 6$\arcmin$) and positioned on the galaxy's bar at a position angle of 145$\degr$. 

Using\,{\sc iraf}\footnote{{\sc iraf} is distributed by the National Optical Astronomy Observatory, which is operated by the Association of Universities for Research in Astronomy (AURA) under cooperative agreement with the National Science Foundation.}, we summed the data in six 10.4$\arcsec$\,$\times$\,4$\arcsec$ zones across the bar, shown in Fig.~\ref{fig:n5430}, and corrected the resulting spectra (3300-8200\,\r{A}) for CCD readout noise, pixel-to-pixel variations and sky brightness. 
Moreover, they have been calibrated in wavelength and flux, respectively with a CuAr calibration lamp and the standard star HD\,109995, and corrected for cosmic rays, dust reddening and redshift. 
The final spectra are shown in Fig.~\ref{spectres}, and have a resolution of 6.1~\r{A}.

Since the first two Balmer lines are measured by our long-slit spectra, we calculated the $E(B-V)$ colour excess \citep{1972ApJ...172..593M,1976AJ.....81..407C} for zones B1 to B6. 
We assumed an optically thick medium, where all the Lyman line photons are scattered to lower energies, as well as an electron temperature of 10\,000~K \citep{2006agna.book.....O}. 
The values derived are given in Table~\ref{tab:result_spectro}, and were applied to the spectra assuming that \cite{1989ApJ...345..245C}'s law holds for NGC\,5430. 
Note that if the H$\beta$ emission is underestimated because of underlying absorption, our $E(B-V)$ values will be overestimated. This is particularly the case for B3, B5 and B6 where the redenning is greater. 
Our uncertainties do not take this effect into account. 

The higher extinction in the B3, B5 and B6 zones suggests that they contain the oldest populations, since the stars have had the time to enrich the interstellar medium with dust.
The lower colour excess in B1 and B2 implies younger populations therein, supported by the presence of a WR bump in the B2 spectrum. 

Note that some giant H\,{\sc ii} regions identified in the SpIOMM data overlap with the zones B1-B6: 19 falls within B2, 25 and 29 fall within B3, 33 and 36 fall within B4, 44 falls within B5, and 45 and 47 fall within B6. 
Even if these zones/regions do not have the same size, their overlap allows for a rough comparison between the long-slit and SpIOMM results, which we discuss in Section~3.

\subsection[]{SpIOMM Flux Validation}

\begin{figure}\centering
\includegraphics[trim = 0cm 0.1cm 0.1cm 0.1cm, clip, width=0.47\textwidth]{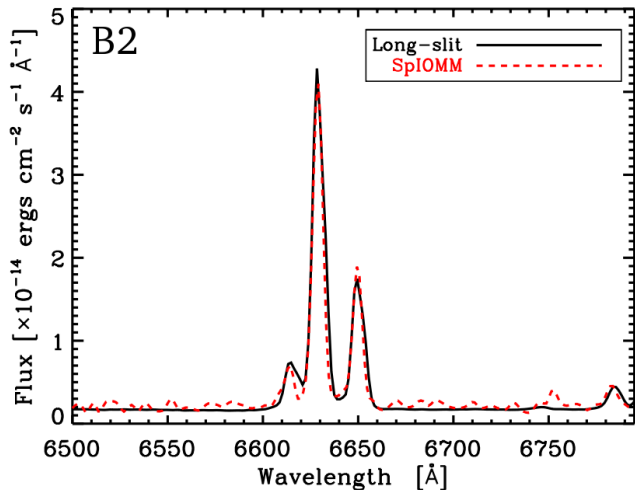}
\caption{Comparison between summed, continuum-subtracted SpIOMM spectrum and long-slit spectrum in zone B2 (see Fig.~\ref{fig:n5430}). Note that the faint feature in the SpIOMM spectrum near 6750~\r{A} is likely an instrumental or processing artifact, rather than real signal.
}\label{comp}
\end{figure}

Since we have carried out the first flux calibration with SpIOMM, we validate the flux scale using the long-slit data. Specifically, we sum the SpIOMM spectra in zone B2 (see Fig.~\ref{fig:n5430}), and compare the result to the corresponding long-slit spectrum. Fig.~\ref{comp} shows the result: there is very good agreement between the normalized SpIOMM and long-slit spectra.

\subsection[]{Measurements and Uncertainties}

Overall, we examine spectra from 51 giant H {\sc ii} regions identified in the SpIOMM data cube and six regions along the bar probed with a long-slit spectrograph. 
All measurements on these spectra were carried out with the\,{\sc iraf} interactive facility {\it splot}.
Gaussian fits were performed on emission lines observed in the SpIOMM spectra (H$\alpha$, [N\,{\sc ii}]\,$\lambda\lambda$6548,6584 and [S\,{\sc ii}]\,$\lambda$6716) and the long-slit spectra (H$\alpha$ and H$\beta$, [OI]$\lambda$6300, [O\,{\sc ii}]$\lambda$3727, [O\,{\sc iii}]$\lambda\lambda$4959,5007, [N\,{\sc ii}]\,$\lambda\lambda$6548,6584 and [S\,{\sc ii}]\,$\lambda\lambda$6716,6731). 
We computed deprojected galactocentric radii for each region using the galaxy disc geometry determined by \cite{2008MNRAS.388..500E}.
The measured H$\alpha$ line fluxes and [N\,{\sc ii}]/H$\alpha$ ratios are shown in Table~\ref{tab:result_spiomm} for the 51 giant H\,{\sc ii} regions detected with SpIOMM, and in Table~\ref{tab:result_spectro} for zones B1-B6.
The uncertainties on our measurements were determined from the continuum variation. 
Note that the emission line fluxes and uncertainties have not been corrected for underlying absorption. However, as found by \cite{Cantin2010}, the difference between the ages of a young population measured in a spectrum that is corrected for the extinction and one that is not is comparable to the uncertainty on the indicator, such as the H$\alpha$, H$\beta$ or  He\,{\sc ii}\,$\lambda$4686 equivalent widths.

\section[]{Results}

Having measured the emission line fluxes in the 51 SpIOMM and six long-slit spectra, we use their ratios to determine the properties of the bar and giant H\,{\sc ii} regions of NGC\,5430.
First, we verify the type of activity (e.g., H\,{\sc ii}, LINER or composite) in these regions with BPT diagrams \citep{1981PASP...93....5B}. 
Then we look at the variation of the [N\,{\sc ii}]$\lambda$6584/H$\alpha$ line ratio with radius.
For regions that exhibit an H\,{\sc ii} region or a composite type of activity, we pursue our investigation by estimating the abundances with the \cite{2002ApJS..142...35K} models. 
Using these abundances to calculate metallicities as well as to constrain the star-formation model {\sc starburst99} \citep{1999ApJS..123....3L}, we estimate the ages of the massive star populations. 
Finally, we compare results obtained with SpIOMM and the long-slit spectrograph as well as with the literature, and investigate the radial dependence of the measured quantities.

\subsection[]{Type of Activity\label{activity}}

\begin{figure}\begin{center}
\includegraphics[trim = 1mm 1mm 0mm 1mm, clip,width=0.475\textwidth]{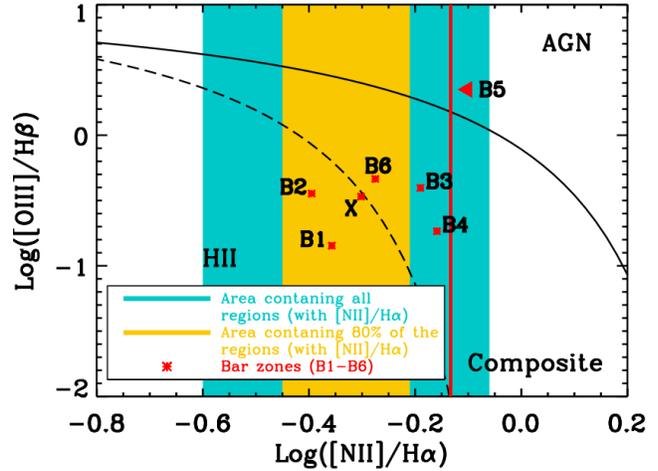}
\caption{Type of activity in the regions and zones. All regions with [N\,{\sc ii}]/H$\alpha$ measurements fall within the blue-green area and 80\% of them in the yellow area. Zone B5 is denoted by the vertical red line, and the other bar zones are shown as symbols. Error bars that are smaller than the symbol size are not plotted.
\label{fig:act_spiomm}}
\end{center} \end{figure}

The first step of our line ratio analysis is to evaluate the type of activity found in the NGC\,5430 bar and to confirm that the 51 detections identified by {\it H{\sevensize\it II}phot.pro} are contained within the H\,{\sc ii} and composite areas of Figure~\ref{fig:act_spiomm}.
For this we used the BPT diagnostic diagram \citep{1981PASP...93....5B,2006MNRAS.372..961K} with the following line ratios: [O\,{\sc iii}]$\lambda$5007/H$\beta$, [N\,{\sc ii}]$\lambda$6584/H$\alpha$, [S\,{\sc ii}]$\lambda$6717,6731/H$\alpha$, [OI]$\lambda$6300/H$\alpha$ and [O\,{\sc iii}]$\lambda$5007/[O\,{\sc ii}]$\lambda$3727. 
All of the requisite line strengths are available from the long-slit data for B1, B2, B3, B4, B6 (and the [O\,{\sc iii}]$\lambda$5007 line is weak in B5), but only [N\,{\sc ii}]/H$\alpha$ is accessible in the SpIOMM spectra.

\begin{figure}
 \begin{center}
\includegraphics[trim = -0.8cm 0cm -0.2cm 0cm, clip, width=0.475\textwidth]{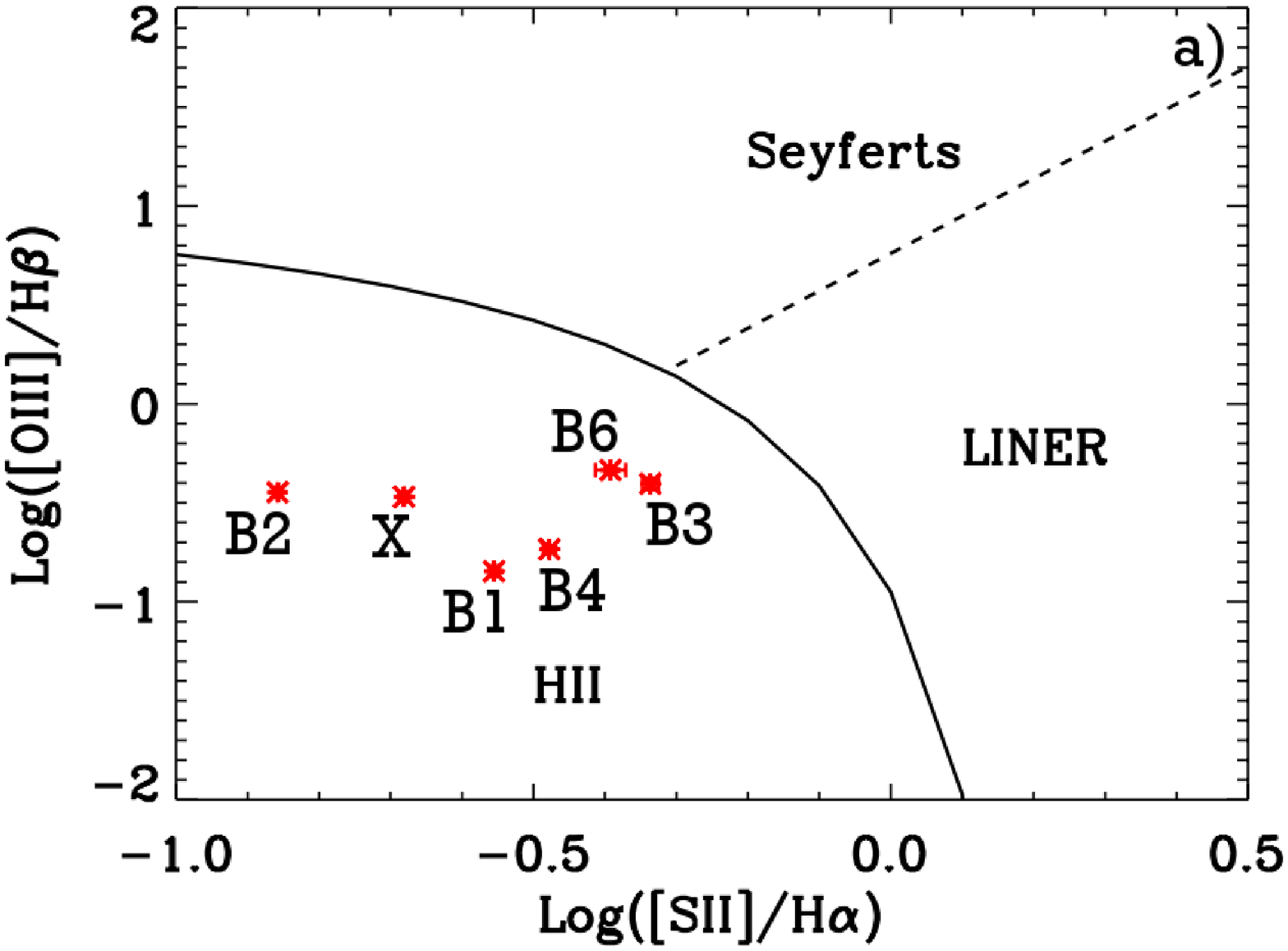}\\\vspace{0.2cm}
\includegraphics[trim = -0.8cm -0.5cm 0cm 0cm, clip, width=0.475\textwidth]{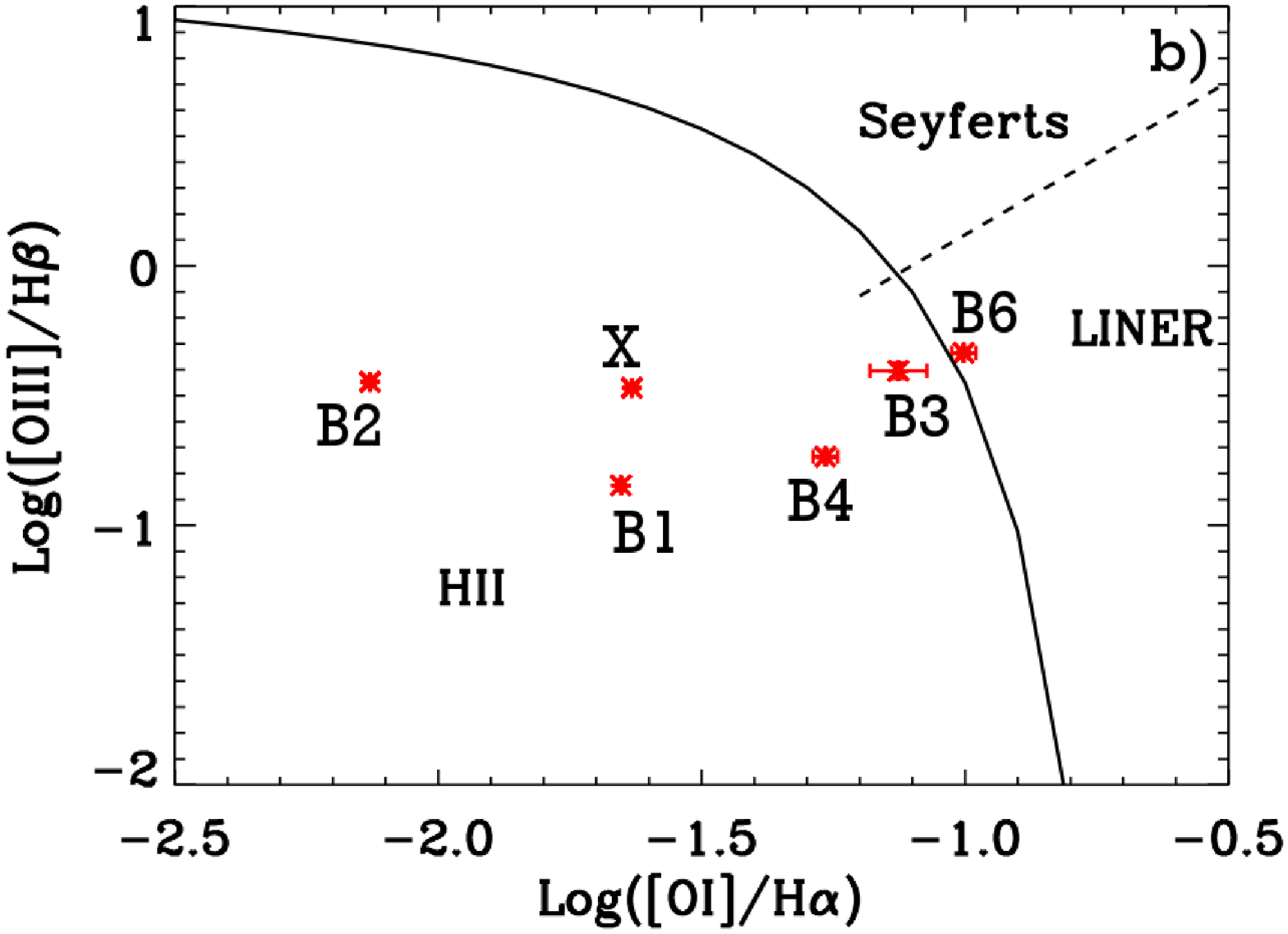}\\\vspace{0.2cm}
\includegraphics[trim = 0cm 0cm -0.2cm 0cm, clip, width=0.475\textwidth]{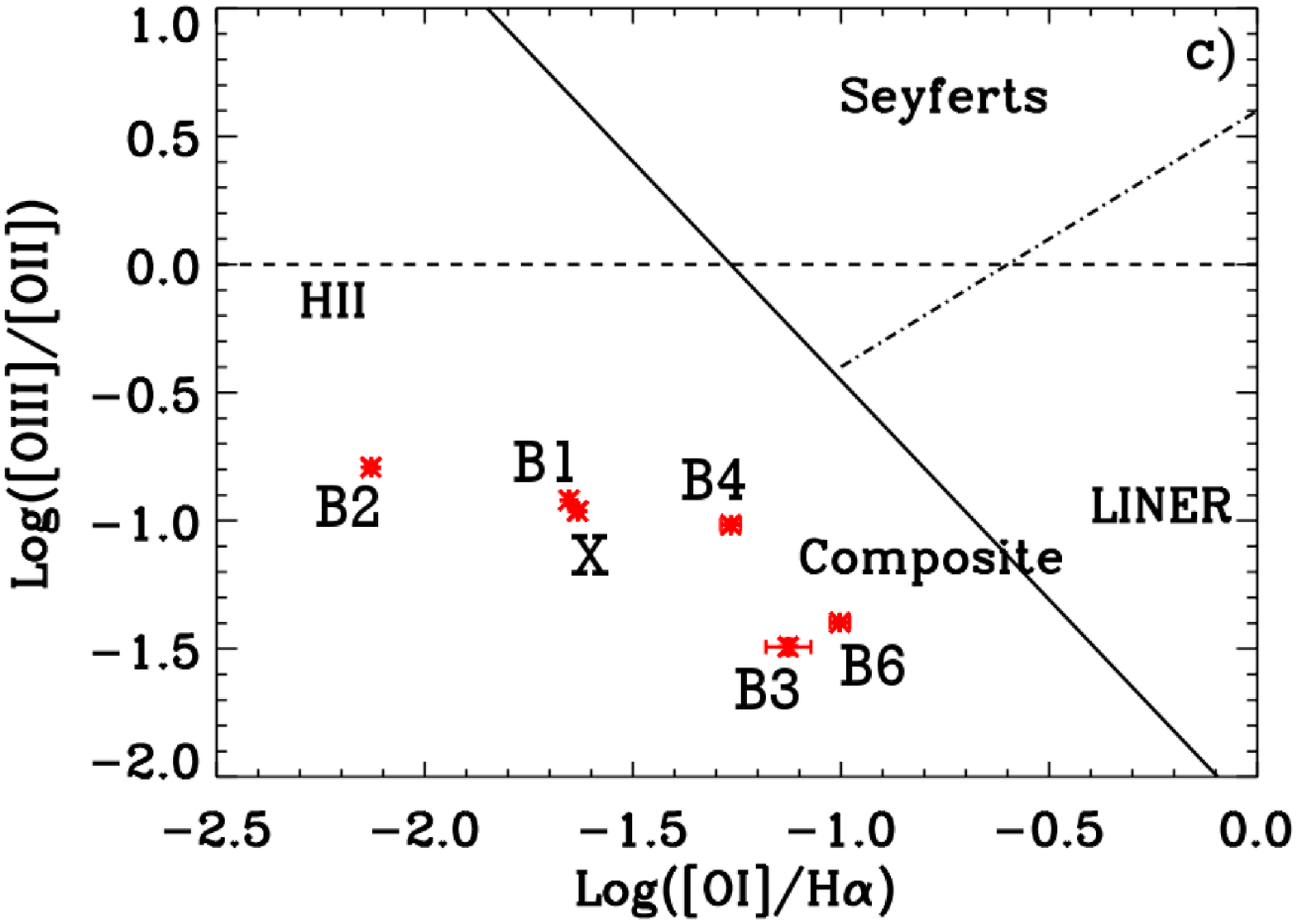}
\caption{
BPT diagram for zones B1-B6, excluding B5, across the bar. The X shows the
ratios obtained for the sum of the emission in B1-B6. Error bars that are smaller than the symbol size are not plotted.
\label{fig:act_spectro}} \end{center} \end{figure}

Fig.~\ref{fig:act_spectro} shows the BPT diagrams for the zones B1, B2, B3, B4 and B6 across the NGC\,5430 bar.
The values from the sum of the spectra across the bar are represented by an X. 
Fig.~\ref{fig:act_spectro} demonstrates two points. 
First, the central region of NGC\,5430 (B4) does not contain a strong AGN, which concurs with the conclusions of \cite{Cantin2010}. 
Second, B2 (WR knot) and B1 (southeast bar end next to the WR knot) show a H\,{\sc ii} type of activity, whereas B3 (between the galaxy centre and the WR knot) and B6 (northwest bar end) are closer to the composite region of the diagram. 
This second result can be explained by non-thermal phenomena, such as compression and shocks, generated by the bar flow \citep{1994ApJ...424..599M}. 

Fig.~\ref{fig:act_spiomm} shows the range of the [N\,{\sc ii}]/H$\alpha$ ratio for 34 of the 51 giant H\,{\sc ii} regions identified in the SpIOMM data with S/N\,$\geq$\,5 in both emission lines. 
The zones B1-B6 (B5 is represented by a vertical red line) are also shown on this BPT diagram.
All the zones/regions are located close to the composite lower limit and concur with the \cite{Cantin2010} results for the central region. 
However, these results do not match our previous ones (Fig.~\ref{fig:act_spectro}) for B1, B2 and the whole bar (X). 
The [N\,{\sc ii}]/H$\alpha$ line ratios presented in Tables~\ref{tab:result_spiomm} and \ref{tab:result_spectro}.  are higher than expected, particularly for the giant H\,{\sc ii} regions that are not located in the bar and are therefore less likely to be affected by nuclear non-thermal process.
B1 and B2 do nonetheless fall into the H\,{\sc ii} region regime of the diagram, consistent with our long-slit results. 
In addition, if we assume that $\log$[O\,{\sc iii}]/H$\beta$\,$<$\,0.5, most of the regions exhibit a H\,{\sc ii} region type of activity and some possibly a composite one, such as regions 12, 22, 24, 31, 40 and 43. 
The non-zero radii of the regions with a composite type of activity suggest that they are not contaminated by a weak central AGN, but more likely by other non-thermal processes. Thus we can proceed with our analysis for the bar and the giant H\,{\sc ii} regions using the \cite{2002ApJS..142...35K} models.

\setcounter{figure}{8}
\begin{center}\begin{figure*}
\includegraphics[trim = 0cm 0cm 1.2cm 0cm, clip, width=0.46\textwidth]{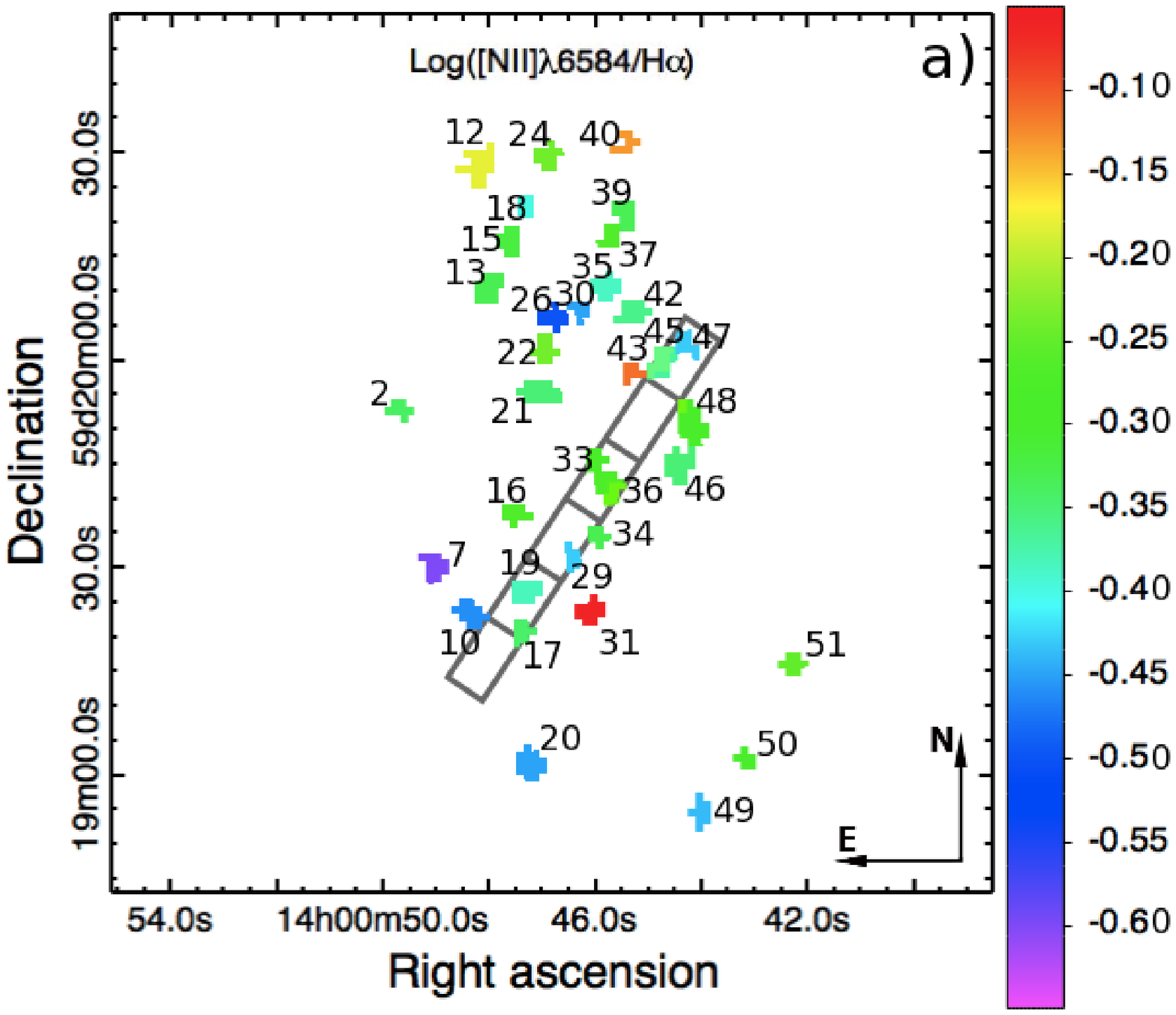}
\includegraphics[trim = -1cm 0cm 0cm 0cm, clip, width=0.53\textwidth]{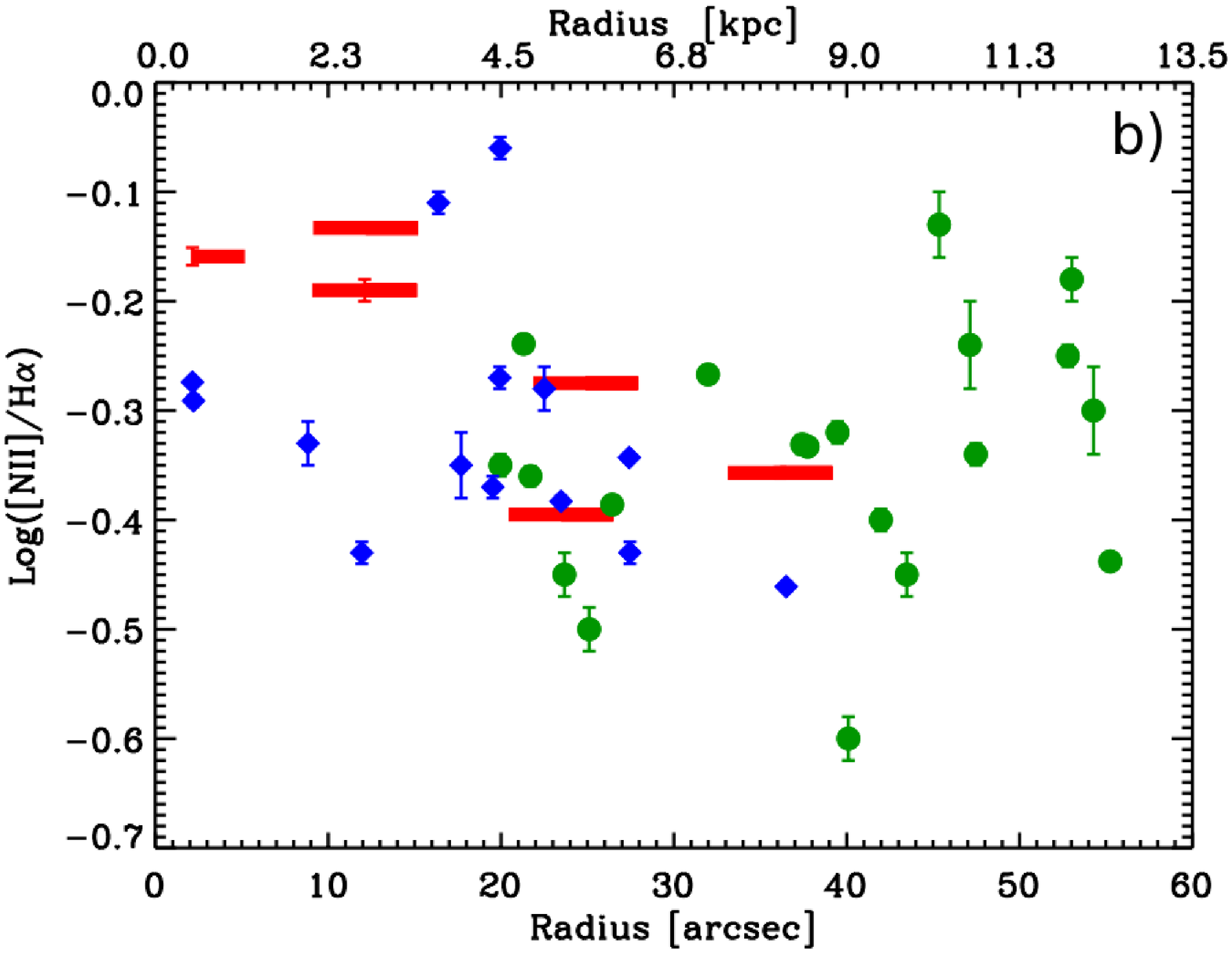}
\includegraphics[trim = 0cm 0cm 0cm 0cm, clip, width=0.46\textwidth]{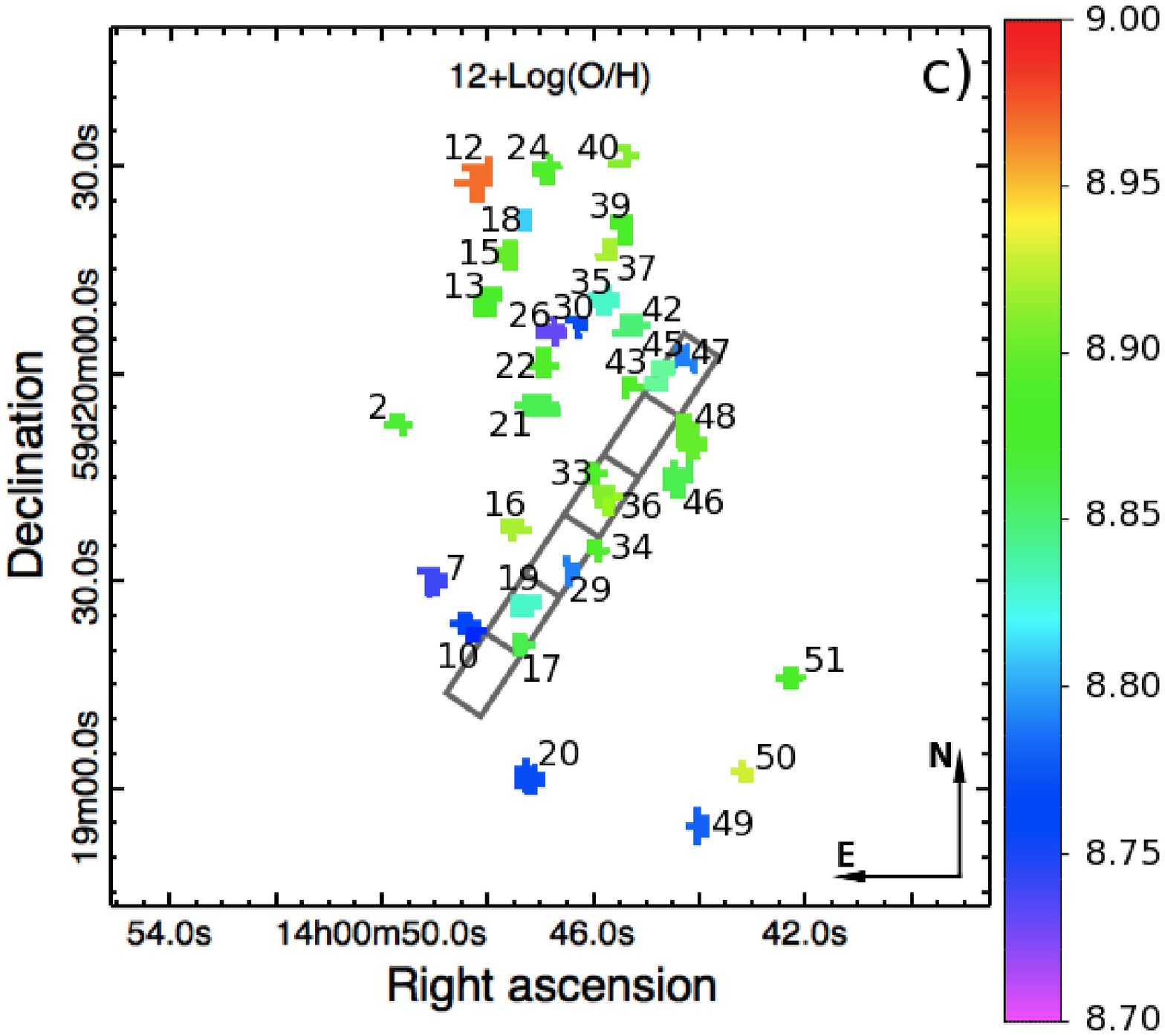}
\includegraphics[trim = -1.55cm 0cm 0cm 0cm, clip, width=0.53\textwidth]{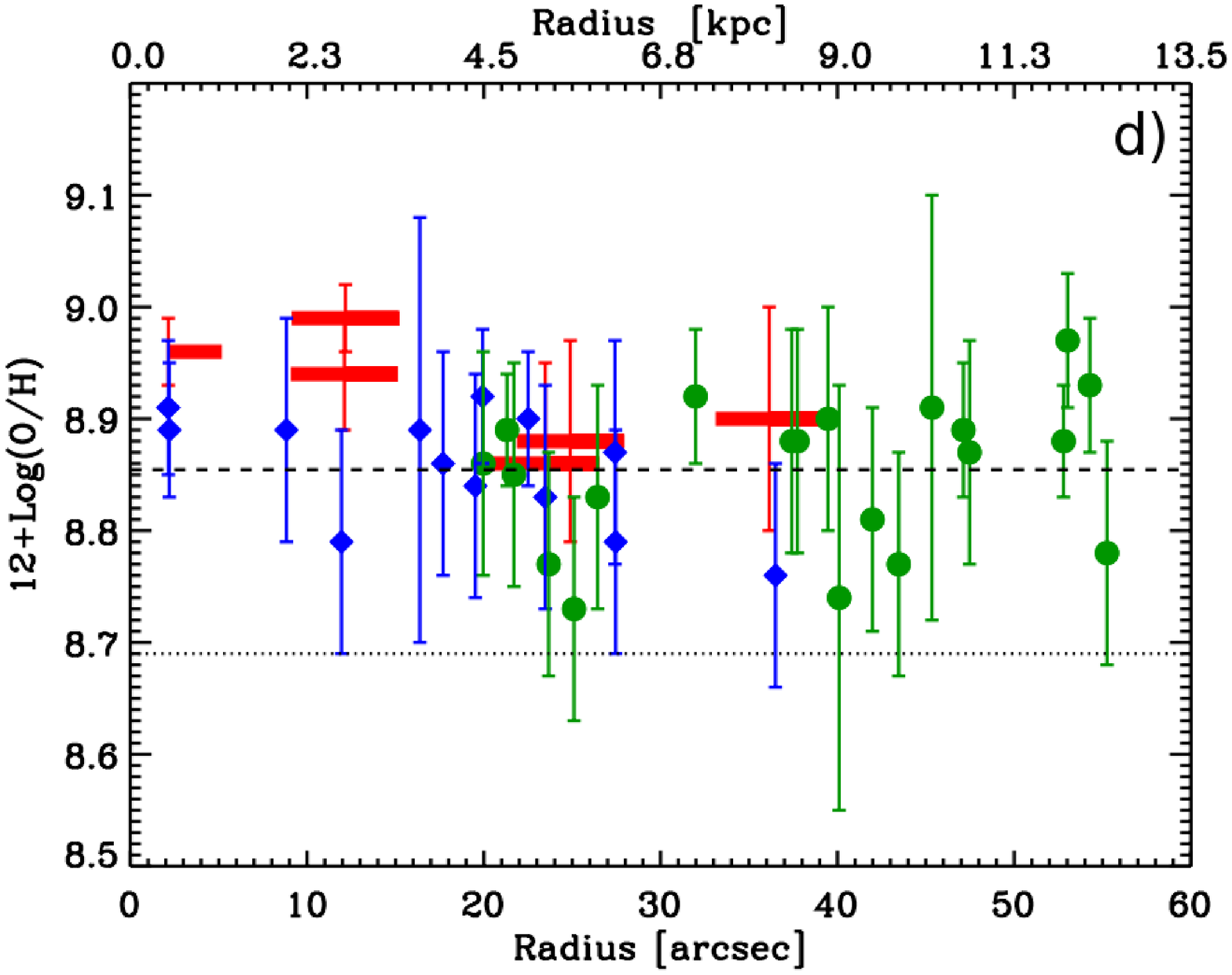}
\includegraphics[trim = 0cm 0cm -0.75cm 0cm, clip, width=0.46\textwidth]{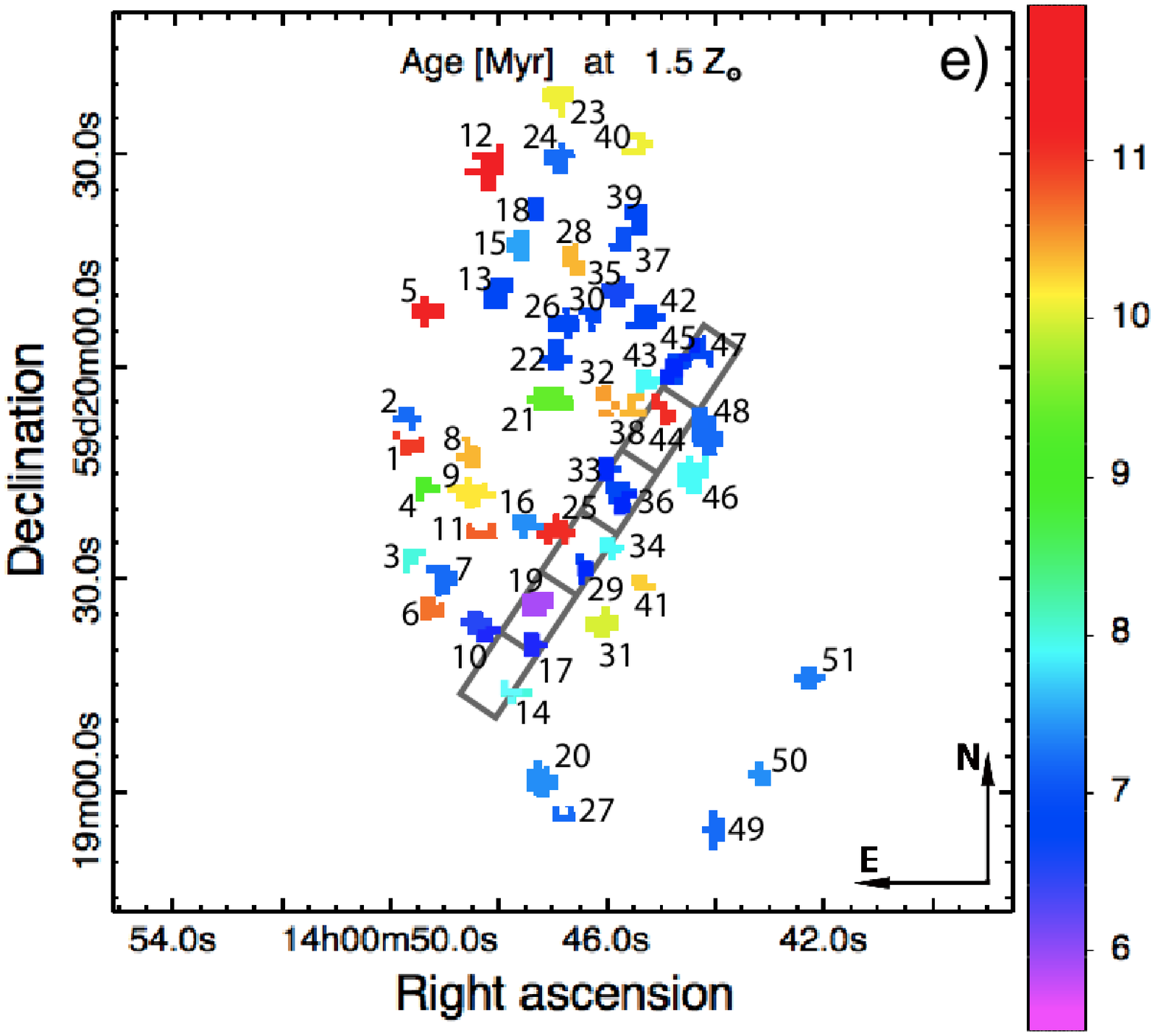}
\includegraphics[trim = -1.75cm 0cm -0.1cm 0cm, clip,  width=0.53\textwidth]{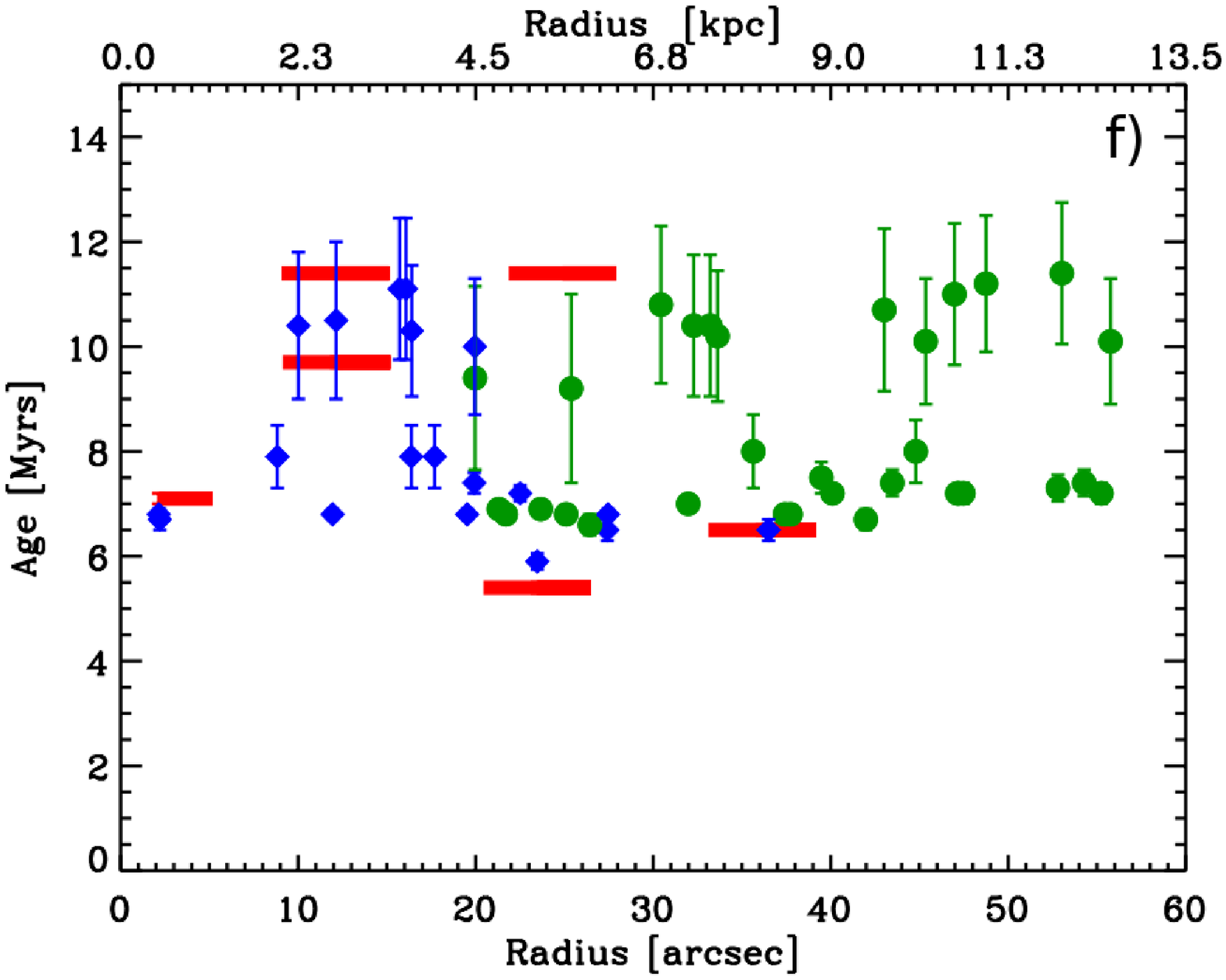}
\caption{Line ratios (top), abundances (middle) and ages (bottom) of giant H\,{\sc ii} regions detected with SpIOMM as well as zones B1-B6. 
The results are presented in the left column as maps and in the right one as a function of galactocentric radius. 
In the left column, the colour scale is shown to the right of each panel and the grey boxes show the locations and sizes of zones B1-B6. 
In the right column, the giant H\,{\sc ii} regions in the bar are shown in blue, those in the arms in green whereas the zones B1-B6 are shown in red. 
The error bars that are smaller than the symbol size are not plotted. 
In d), the dashed and dotted horizontal lines show respectively the measured average oxygen abundance in NGC\,5430 and the solar value \citep{2009ARA&A..47..481A}. 
\label{results}}
\end{figure*}\end{center}

\setcounter{table}{1}
\begin{table*}
\centering
\begin{minipage}{280mm}
\caption{Giant H\,{\sc ii} regions in NGC\,5430.\label{tab:result_spiomm} The number in parentheses denotes the uncertainty in the last digit of the tabulated value.}
\begin{tabular}{@{}lccrrrrrrrrr@{}}
\toprule 
\\
\multirow{2}{*}{No.} 	&\multicolumn{2}{c}{Position (J2000)} 	&\multicolumn{2}{c}{Size}	&\multicolumn{3}{c}{H$\alpha$ line}	&\multirow{2}{*}{$\log$([N\,{\sc ii}]/H$\alpha$)} &\multirow{2}{*}{12+$\log$(O/H)}	&\multicolumn{1}{c}{Z}	& \multicolumn{1}{c}{Age}	 \\\cmidrule(r){2-3}\cmidrule(r){4-5}\cmidrule(r){6-8}

&		$\alpha$	&	$\delta$	&	[pixels]	&	[kpc$^2$] 	&{\tiny $\left[10^{-16}\frac{\rmn{ergs}}{\rmn{cm}^{2}\,\rmn{s}\,\rmn{{\r{A}}}}\right]$} 	&	S/N	&W$_\lambda$~[{\r{A}}]	&	&		&	\multicolumn{1}{c}{[Z$_\rmn{\sun}$]}	&	\multicolumn{1}{c}{[Myr]}	\\

\multicolumn{1}{c}{[1]} & \multicolumn{1}{c}{[2]} & \multicolumn{1}{c}{[3]} & \multicolumn{1}{c}{[4]} & \multicolumn{1}{c}{[5]} & \multicolumn{1}{c}{[6]} & \multicolumn{1}{c}{[7]} & \multicolumn{1}{c}{[8]} & \multicolumn{1}{c}{[9]} & \multicolumn{1}{c}{[10]} &\multicolumn{1}{c}{[11]} & \multicolumn{1}{c}{[12]}\\\hline\\

1	&	       14h00m49.9	&	       59d19m51	&	7	&	1.74	&	        5.1(2)		&	6	&	7.82		&	$--$			&	$--$		&	$--$		&	     11.0(1.3)	\\
2	&	       14h00m49.9	&	       59d19m53	&	7	&	1.74	&	      12.3(2)	&	14	&	20.42	&	$-$0.34(1)	&	8.9(1)	&	 1.5(4)	&	      7.2(2)		\\
3	&	       14h00m49.7	&	       59d19m34	&	6	&	1.49	&	        6.3(2)		&	5	&	14.32	&	$--$			&	$--$		&	 $--$		&	      8.0(6)		\\
4	&	       14h00m49.5	&	       59d19m44	&	6	&	1.49	&	      6.93(8)	&	5	&	13.48	&	$--$			&	$--$		&	 $--$		&	      9.2(1.8	)	\\
5	&	       14h00m49.5	&	       59d20m08	&	10	&	2.48	&	      5.16(5)	&	8	&	6.29		&	$--$			&	$--$		&	 $--$		&	     11.2(1.3)	\\
6	&	       14h00m49.4	&	       59d19m27	&	8	&	1.98	&	        5.9(2)		&	7	&	9.41		&	$--$			&	$--$		&	 $--$		&	     10.7(1.5)	\\
7	&	       14h00m49.3	&	       59d19m32	&	11	&	2.73	&	      25.7(3)	&	21	&	22.49	&	$-$0.60(2)	&	8.7(2)	&	 1.2(2)	&	      7.2(2)		\\
8	&	       14h00m48.7	&	       59d19m47	&	8	&	1.98	&	        7.0(2)		&	6	&	9.85		&	$--$			&	$--$		&	 $--$		&	     10.4(1.3)	\\
9	&	       14h00m48.9	&	       59d19m43	&	13	&	3.22	&	    14.24(4)	&	9	&	10.59	&	$--$			&	$--$		&	 $--$		&	     10.2(1.2)	\\
10	&	       14h00m48.7	&	       59d19m24	&	12	&	2.98	&	    176.9(3)	&	60	&	80.74	&	$-$0.461(4)	&	8.8(1)	&	 1.2(2)	&	      6.5(2)		\\
11	&	       14h00m48.6	&	       59d19m37	&	6	&	1.49	&	        5.6(3)		&	5	&	8.23		&	$--$			&	$--$		&	 $--$		&	     10.8(1.5)	\\
12	&	       14h00m48.6	&	       59d20m28	&	17	&	4.22	&	      7.13(5)	&	7	&	5.62		&	$-$0.18(2	)	&	8.97(6)	&	 1.9(1)	&	     11.4(1.3)	\\
13	&	       14h00m48.2	&	       59d20m11	&	13	&	3.22	&	      43.4(3)	&	17	&	37.44	&	$-$0.331(9)	&	8.9(1)	&	 1.6(4)	&	      6.8(2)		\\
14	&	       14h00m47.9	&	       59d19m15	&	6	&	1.49	&	        6.0(5)		&	6	&	14.63	&	$--$			&	$--$		&	 $--$		&	      8.0(7)		\\
15	&	       14h00m47.8	&	       59d20m18	&	10	&	2.48	&	      15.9(1)	&	12	&	15.88	&	$-$0.32(1)	&	8.9(1)	&	 1.6(5)	&	      7.5(3)		\\
16	&	       14h00m47.7	&	       59d19m39	&	8	&	1.98	&	      18.9(3)	&	9	&	16.4		&	$-$0.27(1)	&	8.92(6)	&	 1.67(9)	&	      7.4(2)		\\
17	&	       14h00m47.5	&	       59d19m22	&	7	&	1.74	&	    110.9(4)	&	66	&	80.01	&	$-$0.343(4)	&	8.9(1)	&	 1.5(4)	&	      6.5(2)		\\
18	&	       14h00m47.4	&	       59d20m23	&	6	&	1.49	&	      27.1(2)	&	33	&	43.28	&	$-$0.40(1)	&	8.8(1)	&	 1.3(3)	&	      6.7(2)		\\
19	&	       14h00m47.5	&	       59d19m28	&	11	&	2.73	&	 2123.0(6)		&	102	&	128.83	&	$-$0.383(5)	&	8.8(1)	&	 1.4(3)	&	      5.9(2)		\\
20	&	       14h00m47.4	&	       59d19m03	&	14	&	3.47	&	      17.2(2)	&	14	&	16.73	&	$-$0.45(2)	&	8.8(1)	&	 1.2(3)	&	      7.4(2)		\\
21	&	       14h00m47.5	&	       59d19m55	&	15	&	3.72	&	      25.3(4)	&	11	&	13.01	&	$-$0.35(1)	&	8.9(1)	&	 1.5(4)	&	      9.4(1.8)	\\
22	&	       14h00m47.2	&	       59d20m01	&	10	&	2.48	&	      33.6(3)	&	25	&	27.93	&	$-$0.239(8)	&	8.89(5)	&	 1.6(1)	&	      6.9(1)		\\
23	&	       14h00m47.2	&	       59d20m39	&	11	&	2.73	&	        8.1(3)		&	7	&	11.06	&	$--$			&	$--$		&	      $--$	&	     10.1(1.2)	\\
24	&	       14h00m47.1	&	       59d20m30	&	10	&	2.48	&	      20.0(6)	&	15	&	22.4		&	$-$0.24(4)	&	8.89(6)	&	     1.6(2)	&	      7.2(2)		\\
25	&	       14h00m47.2	&	       59d19m37	&	11	&	2.73	&	         22(4)		&	7	&	6.91		&	$--$			&	$--$		&	      $--$	&	     11.1(1.3)	\\
26	&	       14h00m47.1	&	       59d20m07	&	10	&	2.48	&	    41.31(4)	&	20	&	30.18	&	$-$0.50(2)	&	8.7(1)	&	     1.1(2)	&	      6.8(1)		\\
27	&	       14h00m47.0	&	       59d18m57	&	5	&	1.24	&	      8.39(7)	&	16	&	22.64	&	$--$			&	$--$		&	      $--$	&	      7.2(2)		\\
28	&	       14h00m46.8	&	       59d20m17	&	9	&	2.23	&	        9.4(3)		&	8	&	9.89		&	$--$			&	$--$		&	      $--$	&	     10.4(1.3)	\\
29	&	       14h00m46.5	&	       59d19m33	&	6	&	1.49	&	      80.9(7)	&	35	&	35.5		&	$-$0.43(1)	&	8.8(1)	&	     1.2(3)	&	      6.8(1)		\\
30	&	       14h00m46.5	&	       59d20m08	&	6	&	1.49	&	      30.4(4)	&	20	&	29.03	&	$-$0.45(2)	&	8.8(1)	&	     1.2(3)	&	      6.9(1)		\\
31	&	       14h00m46.3	&	       59d19m24	&	11	&	2.73	&	      19.2(2)	&	9	&	11.72	&	$-$0.06(1)	&	$--$		&	      $--$	&	     10.0(1.3)	\\
32	&	       14h00m46.1	&	       59d19m57	&	7	&	1.74	&	         17(2)		&	7	&	9.67		&	$--$			&	$--$		&	      $--$	&	     10.5(1.5)	\\
33	&	       14h00m46.1	&	       59d19m47	&	7	&	1.74	&	       442(2)	&	36	&	43.17	&	$-$0.291(5)	&	8.89(6)	&	     1.6(3)	&	      6.7(2)		\\
34	&	       14h00m46.1	&	       59d19m36	&	6	&	1.49	&	      32.5(4)	&	16	&	15.39	&	$-$0.33(2)	&	8.9(1)	&	     1.6(5)	&	      7.9(6)		\\
35	&	       14h00m46.0	&	       59d20m11	&	11	&	2.73	&	      96.4(3)	&	40	&	59.94	&	$-$0.386(2)	&	8.8(1)	&	     1.4(3)	&	      6.6(2)		\\
36	&	       14h00m46.0	&	       59d19m43	&	12	&	2.98	&	     1008(4)	&	40	&	33.43	&	$-$0.274(2)	&	8.91(6)	&	     1.7(4)	&	      6.8(1)		\\
37	&	       14h00m45.9	&	       59d20m17	&	7	&	1.74	&	      21.7(2)	&	17	&	26.63	&	$-$0.267(8)	&	8.92(6)	&	     1.70(8)	&	      7.0(1)		\\
38	&	       14h00m45.7	&	       59d19m54	&	6	&	1.49	&	      19.0(5)	&	7	&	9.76		&	$--$			&	$--$		&	      $--$	&	     10.4(1.4)	\\
39	&	       14h00m45.6	&	       59d20m23	&	10	&	2.48	&	      49.0(3)	&	33	&	41.47	&	$-$0.333(6)	&	8.9(1)	&	     1.5(4)	&	      6.8(2)		\\
40	&	       14h00m45.7	&	       59d20m30	&	7	&	1.74	&	        7.3(3)		&	8	&	11.25	&	$-$0.13(3)	&	8.9(2)	&	     1.7(4)	&	     10.1(1.2)	\\
41	&	       14h00m45.5	&	       59d19m30	&	5	&	1.24	&	        8.0(2)		&	6	&	10.18	&	$--$			&	$--$		&	      $--$	&	     10.3(1.2)	\\
42	&	       14h00m45.6	&	       59d20m06	&	11	&	2.73	&	      58.0(4)	&	23	&	32.75	&	$-$0.360(9)	&	8.9(1)	&	     1.4(4)	&	      6.8(1)		\\
43	&	       14h00m45.4	&	       59d19m59	&	6	&	1.49	&	      26.0(4)	&	14	&	15.26	&	$-$0.11(1)	&	8.9(2)	&	     1.6(4)	&	      7.9(6)		\\
44	&	       14h00m45.1	&	       59d19m56	&	7	&	1.74	&	      20.4(2)	&	8	&	7.36		&	$--$			&	$--$		&	      $--$	&	     11.1(1.3)	\\
45	&	       14h00m45.0	&	       59d19m59	&	12	&	2.98	&	       130(1)	&	30	&	33.59	&	$-$0.37(1)	&	8.8(1)	&	     1.4(4)	&	      6.8(1)		\\
46	&	       14h00m44.7	&	       59d19m46	&	15	&	3.72	&	      46.4(8)	&	19	&	15.02	&	$-$0.35(3)	&	8.9(1)	&	     1.5(5)	&	      7.9(6)		\\
47	&	       14h00m44.4	&	       59d20m04	&	8	&	1.98	&	      75.6(4)	&	31	&	35.4		&	$-$0.43(1)	&	8.8(1)	&	     1.2(3)	&	      6.8(1)		\\
48	&	       14h00m44.4	&	       59d19m54	&	18	&	4.46	&	      91.6(8)	&	20	&	23.48	&	$-$0.28(2)	&	8.9(6	)	&	     1.6(1)	&	      7.2(2)		\\
49	&	       14h00m44.2	&	       59d18m55	&	9	&	2.23	&	   14.69(5)		&	20	&	22.06	&	$-$0.438(7)	&	8.8(1)	&	     1.2(3)	&	      7.2(2)		\\
50	&	       14h00m43.3	&	       59d19m03	&	6	&	1.49	&	        9.0(3)		&	15	&	18.81	&	$-$0.30(4)	&	8.93(6)	&	     1.8(2)	&	      7.4(2)		\\
51	&	       14h00m42.5	&	       59d19m16	&	8	&	1.98	&	      12.6(2)	&	20	&	20.39	&	$-$0.25(1)	&	8.88(5)	&	     1.5(1)	&	      7.3(2)		\\\\\bottomrule\\
\multicolumn{12}{p{0.65\textwidth}}{Note: [1] Region numbers; [2] and [3] Coordinates given by the {\sc idl} program {\it H{\sevensize\it II}phot.pro}; [4] and [5] Sizes of the giant H\,{\sc ii} regions in pixels and in kpc$^2$; [6], [7] and [8] H$\alpha$ emission line fluxes, S/N and equivalent widths W$_\lambda$; [9] $\log$([N\,{\sc ii}]/H$\alpha$) line ratios; [10] oxygen abundances 12+$\log$(O/H) estimated with $\log$([N\,{\sc ii}]/H$\alpha$) ratios; [11] gas metallicities Z estimated with $\log$([N\,{\sc ii}]/H$\alpha$) ratios; [12] Ages of the young star populations estimated with {\sc starburst99} using an average gas metallicity of Z\,=\,1.5\,Z$_\rmn{\sun}$ (Section~\ref{age}) and W$_\lambda$[H$\alpha$].} 
\end{tabular}\end{minipage}\end{table*}

\setcounter{table}{2}
\begin{table*}
 \centering
 \begin{minipage}{180mm}
  \caption{Reddening, metallicity, ionization factor and age of the bar.\label{tab:result_spectro} The number in parentheses denotes the uncertainty in the last digit of the tabulated value.}
\begin{tabular}{@{}l@{\ \ }r@{\ \ }r@{\ }r@{\ \ }r@{\ \ }r@{\ \ }r@{\ \ }r@{\ \ }r@{\ \ }r@{\,}r@{\ \ }r@{\ \ }r@{\ }r@{\ }r@{\ }r@{}}\toprule 
\\
\multirow{2}{*}{No.}	
& \multicolumn{1}{c}{H$\alpha$ flux}&  \multicolumn{1}{c}{\multirow{2}{*}{H$\alpha$/H$\beta$}} & \multicolumn{1}{c}{\multirow{2}{*}{$E(B-V)$}} & \multicolumn{6}{c}{Logarithmic line ratios and oxygen abundance estimates}& \multicolumn{2}{c}{$q$} &\multicolumn{3}{c}{Age ($\pm$0.1) [Myr]}  \\\cmidrule(r){5-10}\cmidrule(r){11-12}\cmidrule(r){13-15}

     &	\tiny	$\left[\frac{10^{-16}\rmn{ergs}}{\rmn{cm}^{2}\,\rmn{s}\,\rmn{{\r{A}}}}\right] $ &&&\multicolumn{1}{c}{[N{\sc ii}]/[O{\sc ii}]}&\multicolumn{1}{c}{O/H}&\multicolumn{1}{c}{[N{\sc ii}]/H$\alpha$} &\multicolumn{1}{c}{O/H}&\multicolumn{1}{c}{$\rmn{R}_{23}$}&\multicolumn{1}{c}{O/H}&\multicolumn{1}{c}{[O{\sc iii}]/[O{\sc ii}]}&\tiny [10$^{6}\frac{\rmn{cm}}{\rmn{s}}$]& \multicolumn{1}{c}{H$\alpha$} & \multicolumn{1}{c}{H$\beta$} & \multicolumn{1}{c}{He{\sc ii}}	 \\
     
\multicolumn{1}{c}{[1]} & \multicolumn{1}{c}{[2]} & \multicolumn{1}{c}{[3]} & \multicolumn{1}{c}{[4]} & \multicolumn{1}{c}{[5]} & \multicolumn{1}{c}{[6]} & \multicolumn{1}{c}{[7]} & \multicolumn{1}{c}{[8]} & \multicolumn{1}{c}{[9]} & \multicolumn{1}{c}{[10]} &\multicolumn{1}{c}{[11]} & \multicolumn{1}{c}{[12]} & \multicolumn{1}{c}{[13]} & \multicolumn{1}{c}{[14]} & \multicolumn{1}{c}{[15]}
     \\\hline\\

B1 &479(1)	&3.932(4)	&0.278(1)	&$-$0.449(6)	&8.93(3)	&$-$0.357(5) 	&8.9(1) 	&0.64(2)	&8.882(9)	&$-$0.920(8)	&20.5(2)	&6.5 	 &6.5	&$--$	\\
B2 & 3979(7)	&3.247(8)	&0.109(1)	&$-$0.278(6)	&9.01(2)	&$-$0.395(3) 	&8.82(9)	&0.43(1)	&9.032(4)	&$-$0.792(9)	&32.5(3)	&5.4		&5.0		&4.9		\\
B3 &982(9)	&6.50(1)	&0.720(8)	&$-$0.78(2)	&8.76(6)	&$-$0.19(1) 	&8.95(5)	&1.1(1)	&$--$			&$-$1.49(3)	&6.8(3) 	&11.4	&12.5	&$--$	\\
B4 &2540(40)	&4.622(8)	&0.420(3)	&0.04(2)		&9.13(2)	&$-$0.159(8)	&8.96(3)	&0.32(1)	&9.048(9)	&$-$1.02(3)	&48.0(6)	&7.1		&7.2		&$--$	\\
B5 &1003(8)	&7.92(1)	&0.89(1)	&$-$0.810(3)	&8.74(5)	&$-$0.133(4)	&8.99(3)	&1.19(3)	&$--$ 			&$--$		&$--$	&9.7		&11.4	&$--$	\\
B6 &521.5(5)	&6.70(2)	&0.75(1)	&$-$0.826(7)	&8.73(5)	&$-$0.275(4)	&8.90(9)	&1.08(2)	&$--$			&$-$1.40(2)	&7.8(2)	&11.4	&12.4	&$--$	\\
X   &12060(30)	&4.191(4)	&0.334(1)	&$-$0.320(7) 	&8.99(3)	&$-$0.302(3) 	&8.93(7)	&0.57(1)	&8.958(7)	&$-$0.96(2)	&22.7(3) 	&6.9		&6.9		&$--$	\\\\\bottomrule\\
\multicolumn{16}{p{1\textwidth}}{Note: [1] Region numbers; [2] H$\alpha$ emission line fluxes; [3] H$\alpha$/H$\beta$ line ratios; [4] internal extinctions $E(B-V)$; [5] $\log$([N\,{\sc ii}]/[O\,{\sc ii}]) line ratios; [6] oxygen abundances 12+$\log$(O/H) estimated from $\log$([N\,{\sc ii}]/[O\,{\sc ii}]) ratios; [7] $\log$([N\,{\sc ii}]/H$\alpha$) line ratios; [8] oxygen abundances 12+$\log$(O/H) estimated from $\log$([N\,{\sc ii}]/H$\alpha$) ratios; [9] R$_{23}$\,=\,([O\,{\sc ii}]+[O\,{\sc iii}])/H$\beta$ line ratios; [10] oxygen abundances 12+$\log$(O/H) estimated from $\log$(R$_{23}$) ratios; [11] $\log$([O\,{\sc iii}]/[O\,{\sc ii}]) line ratios; [12] ionization factors $q$; [13], [14] and [15] Ages of the young star populations estimated with {\sc starburst99} using $\log$(R$_{23}$), $\log$([N\,{\sc ii}]/[O\,{\sc ii}]) or $\log$([N\,{\sc ii}]/H$\alpha$) ratios in order of preference, and the equivalent widths, respectively $\log$(W$_\lambda$[H$\alpha$]), $\log$(W$_\lambda$H$\beta$]) and W$_\lambda$[He {\sc ii}].} 
\end{tabular}\end{minipage}\end{table*}

\subsection[]{[N\,{\sevensize\bf II}]$\lambda$6584/H$\alpha$ line ratio\label{han2}}

As previously mentioned, only the [N\,{\sc ii}]/H$\alpha$ ratio is available for both the SpIOMM and long-slit spectra because of the restricted $r\arcmin$ filter bandwidth used with SpIOMM; the ratios for each zone/region are given in Tables~\ref{tab:result_spiomm} and \ref{tab:result_spectro}.

Fig.~\ref{results} a) and b) show the value of this ratio for regions with S/N\,$\geq$\,5 in both lines, on the sky and as a function of galactocentric radius respectively. 
The logarithmic values range from $-$0.60$\pm$0.02 (region~7)\,$\leq$\,$\log$([N\,{\sc ii}]$\lambda$6584/H$\alpha$)\,$\leq$\,$-$0.06$\pm$0.01 (region~31) with no discernible trend with radius either in the bar or the arms. 
The lack of radial variation in the [N\,{\sc ii}]/H$\alpha$ line ratio, assuming a similar ionization source for all H\,{\sc ii} regions, supports the hypothesis that a chemical mixing mechanism is at work in the galaxy disc. 
We discuss this further in Section~5.

We also compared the long-slit and SpIOMM results in cases where the {\it H{\sevensize\it II}phot.pro} detections overlapped with the zones B1-B6 (see Section \ref{GHIIR}). 
Note that the long-slit zones are at least two times larger than the SpIOMM regions. 
We found $\log$([N\,{\sc ii}]/H$\alpha$)\,=\,$-$0.395$\pm$0.003 for B2 (long-slit) and $\log$([N\,{\sc ii}]/H$\alpha$)\,=\,$-$0.383$\pm$0.005 for 19 (SpIOMM), values that are consistent at the two sigma level despite the different sizes of these regions. 
This is not surprising, because the emission in B2 and 19 is likely dominated by the WR knot consistent with the results of Fig.~\ref{fig:act_spectro}.
However, near the galaxy centre, B3 ($\log$([N\,{\sc ii}]/H$\alpha$)\,=\,$-$0.19$\pm$0.01), B4 ($\log$([N\,{\sc ii}]/H$\alpha$)\,=\,$-$0.159$\pm$0.008) and B5 ($\log$([N\,{\sc ii}]/H$\alpha$)\,=\,$-$0.133$\pm$0.004) have higher ratios than their corresponding giant H\,{\sc ii} regions, i.e. 29 ($\log$([N\,{\sc ii}]/H$\alpha$)\,=\,$-$0.43$\pm$0.01), 33 ($\log$([N\,{\sc ii}]/H$\alpha$)\,=\,$-$0.291$\pm$0.005) and 36 ($\log$([N\,{\sc ii}]/H$\alpha$)\,=\,$-$0.274$\pm$0.002) respectively. 
Since H\,{\sc ii} regions tend to have a lower [N\,{\sc ii}]/H$\alpha$ ratio \citep{1981PASP...93....5B} than their surroundings and the long-slit zones are not limited to these structures, this discrepancy is also to be expected. 
It is corroborated by an OASIS (Optically Adaptive System for Imaging Spectroscopy) map from \cite{Cantin2010} of the NGC\,5430 central region, on a scale slightly smaller that of B4. 
The same phenomenon occurs with B6 ($\log$([N\,{\sc ii}]/H$\alpha$)\,=\,$-$0.275$\pm$0.004) and its associated SpIOMM regions 45 ($\log$([N\,{\sc ii}]/H$\alpha$)\,=\,$-$0.37$\pm$0.01) and 47 ($\log$([N\,{\sc ii}]/H$\alpha$)\,=\,$-$0.43$\pm$0.01).

\subsection[]{Oxygen Abundance, Ionization Factor and Metallicity\label{abundances}}

With the type of activity in the NGC\,5430 bar and giant H\,{\sc ii} regions established, we estimate oxygen abundances  12+$\log$(O/H). 
Since the abundance and ionization factor are interdependent, we employ the iterative method proposed by \cite{2002ApJS..142...35K} that uses strong optical emission line ratios and offers alternatives if one is missing. 
For our study, in the first step of this technique where a rough estimate of the abundance is obtained, we use [N\,{\sc ii}]$\lambda$6584/[O\,{\sc ii}]$\lambda$3727 or [N\,{\sc ii}]$\lambda$6584/H$\alpha$ in order of preference. 
As explained in Section~\ref{activity}, the second of these ratios is most sensitive to non-thermal process and thus the least desirable.
Then, with this first abundance estimate and the [O\,{\sc iii}]$\lambda$5007/[O\,{\sc ii}]$\lambda$3727 ratio (if available), we estimate the ionization factor. 
In the third and last step of this iterative method, a more precise abundance estimate is obtained with the R$_{23}$\,=\,([O\,{\sc ii}]+[O\,{\sc iii}])/H$\beta$ ratio. 
Finally, if the abundance estimates obtained from the first and third steps do not agree with each other, a new iteration is required. 

It was possible to apply the method to the B1, B2 and B4 spectra. 
The underlying physics in B3 and B6 are not covered by the R$_{23}$ model,  leaving us with rough abundance and ionization factor estimates for these two zones. 
Recall that the [O\,{\sc iii}]$\lambda$5007 line cannot be measured in B5 or in the SpIOMM spectra, so only rough abundance estimates were obtained via step~1 described above. 
Moreover, the model does not apply to region~31, because of its elevated [N\,{\sc ii}]/H$\alpha$ ratio. 
Overall, we obtained at least an abundance estimate for B1-B6 and for 33 of the 51 giant H\,{\sc ii} zones/regions with the [N\,{\sc ii}]/H$\alpha$ ratio (S/N\,$\ge$\,5 in both lines), summarized in Tables~\ref{tab:result_spiomm} and \ref{tab:result_spectro}.
Our uncertainties are limited to those of the models, except for the abundances estimated from the R$_{23}$ ratio, where measurement uncertainties dominate.

First, we compared the abundances estimated from the line ratios available for B1-B6 and the whole bar (X) (see Table~\ref{tab:result_spectro}). 
For B1, B2, B3, B6 and the whole bar (X), all the calculated values are consistent within each zone at the three sigma level while the estimates for B4 and B5, the central region of NGC\,5430 and that next to it, are not. 
According to \cite{Cantin2010} and in agreement with the results of Section~\ref{activity}, the central region of NGC\,5430 contains a weak AGN, and therefore non-thermal processes influence the emission line ratios. 

As expected from the intensity maps in Section~\ref{intensity}, the strongest ionization factors $q$ are measured in B4 ($q$\,=\,48.0$\pm$0.6\,$\times$10$^6$\,cm\,s$^{-1}$), then B2 ($q$\,=\,32.5$\pm$0.3\,$\times$10$^6$\,cm\,s$^{-1}$), B1 ($q$\,=\,20.5$\pm$0.2\,$\times$10$^6$\,cm\,s$^{-1}$), B6 ($q$\,=\,7.8$\pm$0.2\,$\times$10$^6$\,cm\,s$^{-1}$) and B3 ($q$\,=\,6.8$\pm$0.3\,$\times$10$^6$\,cm\,s$^{-1}$). 
Without a strong enough [O\,{\sc iii}]$\lambda$5007 line, it is not possible to estimate the ionization factor for B5. However, because R$_{23}$ for this zone resembles that of B3 and B6, we expect a relatively low $q$ in B5 as well.
With $q$\,=\,22.7$\pm$3\,$\times$10$^6$\,cm\,s$^{-1}$,  the integrated emission of the bar (X) is clearly dominated by that of its brightest zones B2 and B4.

Fig.~\ref{results} c) and d) show the abundances estimated from [N\,{\sc ii}]/H$\alpha$ line ratios. 
Here again, there is no noticeable gradient nor difference between the values measured in the arms and bar of NGC\,5430. 
The abundances range from 8.7$\pm$0.1 (region~26)\,$\leq$  12+$\log$(O/H)\,$\leq$\,8.97$\pm$0.06 (region~12). 
The average value 12+$\log$(O/H) = 8.86$\pm$0.07 is super-solar ([12+$\log$(O/H)]$_\rmn{\sun}$\,=\,8.69$\pm$0.06, \citealt{2009ARA&A..47..481A}). 
Consistent with the [N\,{\sc ii}]/H$\alpha$ ratios (Section~\ref{han2}) and because B1-B6 include more than just the H\,{\sc ii} regions that fall within them, the average abundances in the former are three standard deviations higher than those in the latter. 
However, an OASIS abundance map of the central region by \cite{Cantin2010} shows higher values (8.92\,$\leq$\,12+$\log$(O/H)\,$\leq$\,9.18) than found here, but they also are homogeneous. 
This discrepancy can be explained by a difference in methodologies. 

Having abundances in-hand estimated from the [N\,{\sc ii}]/H$\alpha$ ratio, we calculated the metallicities of the young star populations (2\,$\leq$\,T\,[Myr]\,$\leq$\,14) of each region using the \cite{2009ARA&A..47..481A} constants. 
We found a range from 1.1$\pm$0.2\,$\leq$\,Z\,[Z$_\rmn{\sun}$]$\leq$\,1.9$\pm$0.1, with an average value of Z\,=\,1.5$\pm$0.2\,Z$_\rmn{\sun}$.

\subsection[]{Young Star Population Age\label{age}}
 
To estimate the age of the young population, we used the {\sc starburst99} star formation model \citep{1999ApJS..123....3L}, with a standard initial mass function (slope $\alpha$\,=\,2.5 and cutoff M$_{\rmn{up}}$\,=\,100\,M$_\rmn{\sun}$) in the case of instantaneous star formation. 
This Web-based software uses 5 different metallicites (0.05\,Z$_\rmn{\sun}$, 0.2\,Z$_\rmn{\sun}$, 0.4\,Z$_\rmn{\sun}$, 1\,Z$_\rmn{\sun}$ or 2\,Z$_\rmn{\sun}$) and the equivalent width (W$_\lambda$) of some emission lines (eg. H$\alpha$ and H$\beta$) as constraints. 
Note that we did not subtract the contamination from old populations. 

First, we evaluated the age of the bar with our best metallicity estimates determined from the R$_{23}$ or [N\,{\sc ii}]/[O\,{\sc ii}] ratio, and the H$\alpha$, H$\beta$ and  He\,{\sc ii}\,$\lambda$4686 equivalent widths when available. 
Since the equivalent width uncertainties are smaller than those of the {\sc starburst99}
model time steps, we adopt the latter (0.1~Myr) as our uncertainty estimate on the stellar population ages. 
The results are presented in Tables~\ref{tab:result_spiomm} and \ref{tab:result_spectro}. 
For the zones with strong emission lines and therefore an abundance estimated from the R$_{23}$ ratio, i.e. B1, B2, B4 and the whole bar (X), the ages found with the H$\alpha$ and H$\beta$ equivalent widths are consistent with each other within three sigma. 
Even if statistically speaking the values found for B3, B5 and B6 do not concur, they remain close. 
Similarly, the age found for B2 using He\,{\sc ii}\,$\lambda$4686 differs from that found using H$\alpha$ by $\sim$3.5 sigma.

Second, we observed in Section~\ref{abundances} that the metallicity estimated from the [N\,{\sc ii}]/H$\alpha$ ratio for the NGC\,5430 bar and arms are homogeneous with a range 1.1$\pm$0.2\,$\leq$\,Z\,[Z$_\rmn{\sun}$]\,$\leq$\,1.9$\pm$0.1 and an average value of 1.5$\pm$0.2\,Z$_\rmn{\sun}$. 
Because this is a rough estimate due to the sensitivity of the [N\,{\sc ii}]/H$\alpha$ ratio to non-thermal processes, we decided to evaluate the age of the young star populations at 1\,Z$_\rmn{\sun}$ and 2\,Z$_\rmn{\sun}$ with the H$\alpha$ equivalent widths. 
We adopt the average of the two results as our age estimate at Z\,=\,1.5\,Z$_\rmn{\sun}$, and half of their difference as our uncertainty on this value. 
Note that even if only 33 of the 51 giant H\,{\sc ii} regions have a metallicity estimate because of S/N or model limitations (see Sections~\ref{han2} and \ref{abundances}), we assumed that all 51 giant H\,{\sc ii} regions have the same metallicity of Z\,=\,1.5\,Z$_\rmn{\sun}$, and computed ages accordingly. 
For transparency, the detailed results are compiled in Tables~\ref{tab:result_spiomm} and~\ref{tab:result_spectro}. 

Because the young stellar populations are located primarily in the giant H\,{\sc ii} regions, we find that the age obtained for B4 (T\,=\,7.2$\pm$0.1\,Myr), and the regions 33 (T\,=\,6.7$\pm$0.2\,Myr) and 36 (T\,=\,6.8$\pm$0.1\,Myr) within B4 agree at the three sigma level. 
These results agree with the \citeauthor{1997A&A...324...41C} (1997; 6$\pm$1\,Myr) and \citeauthor{Cantin2010} (2010; 6.2$\pm$0.8\,Myr in the two central knots) studies. 
As expected, the WR knot is the youngest region in the galaxy, with an age between 4.9$\pm$0.1\,Myr (zone~B2 from W$_\lambda$[He\,{\sc ii}]) and 5.9$\pm$0.2\,Myr (region~19 from $\log$(W$_\lambda$[H$\alpha$])). 
However, if we consider only the values estimated from the H$\alpha$ equivalent with, the lower limit is 5.4$\pm$0.1\,Myr. \citealt{1997A&A...324...41C} estimated an age between 3.4\,$\leq$\,T\,[Myr]\,$\leq$\,6.0 for this structure, which includes all our values.
We have therefore narrowed the age intervals previously determined for both the central region  (6.5\,$\leq$T\,[Myr]\,$\leq$\,6.9) and the WR knot (4.8\,$\leq$T\,[Myr]\,$\leq$\,6.1).

Fig.~\ref{results} e) and f) show the measured young population ages on the sky and as a function of radius, respectively. 
If we focus on the ages estimated from the H$\alpha$ equivalent width, the values range from 5.4$\pm$0.1\,Myr (zone~B2) to 11.4$\pm$1.3\,Myr (region~12). 
However, even taking into account the possible variation of the metallicity between 1\,Z$_\rmn{\sun}$ and 2\,Z$_\rmn{\sun}$, two average ages seem to dominate the young star populations without any discernible trend as a function of radius in either the bar or the arms of NGC\,5430. 
A first wave of massive stars would have been formed in NGC\,5430 10.5$\pm$0.6\,Myr ago (for 18 of the 51 giant H\,{\sc ii} regions) and a second one about 3.4\,Myr later, i.e. 7.1$\pm$0.5\,Myr ago (for 33 of the 51 giant H\,{\sc ii} regions). 

\section[]{Discussion\label{discussion}}

The emission line intensities, oxygen abundances and stellar population ages for the giant H\,{\sc ii} regions and  the bar of NGC\,5430, acquired with the SpIOMM imaging Fourier transform spectrograph and a long-slit spectrograph at the Observatoire du Mont-M\'egantic, are broadly consistent with each other as well as with results from the literature (\citealt{1982PASP...94..765K}, \citealt{1987A&A...172...43K}, \citealt{1996ASPC..103..175C}, \citealt{1997A&A...324...41C}, \citealt{2005A&A...439..539O}, and \citealt{Cantin2010}; see Section~3).

Using {\it H{\sevensize\it II}phot.pro} with a SpIOMM intensity map summed over both H$\alpha$ and [N\,{\sc ii}], we detected 99 bright regions, of which 51 had S/N\,$\geq$\,5 in the H$\alpha$ line. 
Additionally, 13/99 regions had 3\,$\leq$\,S/N\,$<$\,5; an SpIOMM integration three times longer than that obtained would thus have improved our number statistics by about 25\%. 

We estimated the type of activity from BPT diagrams, and found that our SpIOMM detections are likely giant H\,{\sc ii} regions (Section~\ref{activity}). The central region of NGC\,5430 exhibits a composite type of activity, which is consistent with the conclusion of \cite{Cantin2010} that this galaxy does not harbour a strong AGN. 
In addition, by comparing the different BPT diagrams (Sections~\ref{activity} and \ref{han2}), we found that the [N~{\sc ii}]/H$\alpha$ line ratios measured in some regions with strong emission were higher than expected. 
This also been observed by \cite{2011MNRAS.415.2439R} in the spiral galaxy NGC\,628.

We compute the oxygen abundances and young stellar population ages in the giant H\,{\sc ii} regions and in the bar of NGC\,5430 (Fig.~\ref{results}), and find that neither varies 1) between the bar and the spiral arms, nor 2) as a function of galactocentric radius. These results suggest the existence of a mixing mechanism in the disc. 
An obvious candidate for abundance mixing is NGC\,5430's bar, that is thought to transport gas through the disc and mix the chemical elements that it contains. 
The mechanism would lead to a flattening of the abundance gradient as observed by \cite{1981A&A...101..377A}, \cite{1992PhDT........91M}, \cite{1994ApJ...424..599M}, \cite{1996ASPC...91...63R}, \cite{1996ApJ...462..114N} and \cite{2011MNRAS.415.2439R} among others (see \citealt{2011arXiv1101.1771V} for a review).
Thus, the bar would have already mixed the disc in NGC\,5430.
However, since NGC\,5430 is a SB(s)b, our results are also consistent with the observations of \cite{1999ApJ...516...62D} that suggest that the abundance gradient in early-type spiral galaxies, which for them includes S0 to Sb types,  is flatter even in the absence of a bar. 

The positions of the young star populations in the NGC\,5430 disc are consistent with the \cite{2007A&A...465L...1W} simulations. 
The youngest stars of the bar are located in the central region and at the ends, while older stars are found near the centre, perpendicular to the bar. 
Because of their short lifetimes, the young star populations in a galaxy are least affected by scattering in the disc and tend to stay closest to their birthplace. 
If we apply this principle to our observations, it means that a first galaxy-wide wave of star formation was triggered 10.5$\pm$0.6~Myr years ago and a second one 7.1$\pm$0.5~Myr ago.

Both the widespread star formation episodes and the relatively short interval between them seem hard to explain using the dynamics of the bar in NGC\,5430. 
Even if we assume that this galaxy harbours a fast bar where corotation occurs roughly at the bar edge, it would take material at this radius $\sim$100\,Myr to complete one revolution. 
Another hypothesis is that the waves of star formation were triggered by the inflow of gas toward the galaxy centre along the bar. 
However, the regular rotation in the velocity field of NGC\,5430 \citep{2008MNRAS.388..500E} implies that the inflow velocities are unlikely to be larger than $\sim$50\,km\,s$^{-1}$ \citep{2004ApJ...605..183W,2007ApJ...664..204S}, also too slow to explain the existence of two galaxy-wide bursts. Thus while the bar provides a potential mixing mechanism in the disc, it is not clear exactly how the rapid mixing required by our observations would take place.

It is also possible the two waves of star formation detected in NGC\,5430 represent an episode of cyclical star formation in the disc. If the star formation is indeed periodic, we would expect to see a stellar population with an age of $\sim$3.7\,Myr, as well as one aged by $\sim$14\,Myr. The former might be difficult to detect in the optical, while the latter would not ionize the surrounding medium and produce emission lines; it is therefore not inconceivable that they are missed by our data. Note that if NGC\,5430 contains a young, dust-enshrouded star population, then it should be bright in the infrared, as is the case \citep{1987A&A...172...43K,1995ApJS...98..129K,1995ApJS...98..171V}. 

Finally, a high-resolution dynamical study of NGC\,5430 could provide valuable information to help understand the mechanisms driving the evolution of this galaxy, and particularly those that led to the two waves of young star formation reported here.

\section[]{Conclusion\label{conclusion}}

An SpIOMM hyperspectral data cube as well as long-slit spectra, both obtained at the Observatoire du Mont-M\'egantic, have been used to study the properties of the bar and the giant H\,{\sc ii} regions in the spiral galaxy NGC\,5430. 
For the first time, SpIOMM data have been successfully calibrated in flux and cross-compared with a long-slit spectrum, itself independently calibrated with the spectrum of a standard star. 
At each step of our analysis, we compared the results obtained from SpIOMM data to those obtained with long-slit spectrograph data. 
They are broadly consistent with each other, as well as with results published in the literature. 

Exploiting the SpIOMM imaging capacity, we produced H$\alpha$ and [N\,{\sc ii}]$\lambda$6584 intensity maps, from which we identified 51 giant H\,{\sc ii} regions, with S/N\,$\geq$\,5 in H$\alpha$, using the {\sc idl} program {\it H{\sevensize\it II}phot.pro}. 
One spectrum has been obtained for each of these regions. 
Additionally, spectra for six zones across the bar were extracted from the long-slit data.

Using the emission line fluxes measured in these spectra, we evaluated the type of activity, the [N\,{\sc ii}]/H$\alpha$ ratio, the oxygen abundance and the age of the young star populations (2\,$\leq$\,T\,[Myr]\,$\leq$\,14) in every region. 
With BPT diagrams, we confirmed the absence of a strong AGN in the central region of NGC\,5430 and the pure H\,{\sc ii} region nature of the Wolf-Rayet knot.
No discernible variation in the [N\,{\sc ii}]/H$\alpha$ ratio or the oxygen abundance was measured between the bar and spiral arms in NGC\,5430, or as a function of galactocentric radius.
These results are consistent with the hypothesis that a chemical mixing mechanism, possibly the bar, is at work in the galaxy disc.

We find evidence for two distinct young populations of stars in NGC\,5430, one with an age of 7.1$\pm$0.5\,Myr and the second with an age of 10.5$\pm$0.6\,Myr. 
As with the line ratios and abundances, we find no evidence for a radial gradient in either population. 
The short lifetimes of these populations suggest that mixing occurred very rapidly in the disc. While the presence of the bar provides a mixing mechanism, it is unclear how this dynamical structure mixed the disc on such short timescales.

\section*{Acknowledgments}

We thank Anne-Pier Bernier and Maxime Charlebois for their support with the imaging Fourier transform spectrograph SpIOMM during the observations at the Observatoire du Mont-M\'egantic and throughout this project. 
We also thank Carmelle Robert, Laurent Drissen and Rachel Kuzio de Naray for helpful comments on early version of this manuscript.
Finally, we thank the reviewer, John Beckman, for constructive comments that helped to improve the clarity of the paper.

É. B. acknowledges funding from Fonds Québécois de la Recherche sur la Nature et les Technologies (FQRNT) of the Government of Québec and the Hubert-Reeves Fellowship. 
K. S. acknowledges funding from National Sciences and Engineering Reseach Council of Canada (NSERC). 

This research has made use of the NASA/IPAC Extragalactic Database (NED) which is operated by the Jet Propulsion Laboratory, California Institute of Technology, under contract with the National Aeronautics and Space Administration. We acknowledge the usage of the HyperLeda database ({http://leda.univ-lyon1.fr}).

\bibliographystyle{mn2e}
\bibliography{biblio}

\end{document}